\documentclass[a4paper, 11pt]{article}
\usepackage{geometry}
\usepackage[utf8]{inputenc}
\usepackage{amsmath}
\usepackage{amssymb}
\usepackage{natbib}
\usepackage{authblk}
\usepackage{footnote}
\usepackage{graphicx}
\usepackage{tikz}
\usepackage{bm}
\usepackage{subcaption}
\usepackage{xcolor} 
\usepackage{url}
\usepackage[colorlinks=true,citecolor=blue,linkcolor=blue,urlcolor=blue]{hyperref}
\usepackage{booktabs}
\usepackage{array}
\usepackage[bottom]{footmisc}
\usepackage[ruled, vlined]{algorithm2e}
\usepackage{bbm}

\usepackage{setspace}
\usepackage{lscape}
\usepackage{lineno}

\title{A changepoint approach to modelling non-stationary soil moisture dynamics}

\author[1]{Mengyi Gong \thanks{Corresponding address: Department of Mathematics and Statistics, Lancaster University, Lancaster, U.K., LA1 4YF, E-mail: \url{m.gong1@lancaster.ac.uk}}}
\author[1]{Rebecca Killick}
\author[1]{Christopher Nemeth}
\author[2]{John Quinton}
\affil[1]{Department of Mathematics and Statistics, Lancaster University, Lancaster, U.K.} 
\affil[2]{Lancaster Environment Centre, Lancaster University, Lancaster, U.K.} 

\date{}

\begin{document}

\maketitle

\abstract{Soil moisture dynamics provide an indicator of soil health that scientists model via drydown curves. The typical modelling process requires the soil moisture time series to be manually separated into drydown segments and then exponential decay models are fitted to them independently. Sensor development over recent years means that experiments that were previously conducted over a few field campaigns can now be scaled to months or years at a higher sampling rate. To better meet the challenge of increasing data size, this paper proposes a novel changepoint-based approach to automatically identify structural changes in the soil drying process and simultaneously estimate the drydown parameters that are of interest to soil scientists. A simulation study is carried out to demonstrate the performance of the method in detecting changes and retrieving model parameters. Practical aspects of the method such as adding covariates and penalty learning are discussed. The method is applied to hourly soil moisture time series from the NEON data portal to investigate the temporal dynamics of soil moisture drydown. We recover known relationships previously identified manually, alongside delivering new insights into the temporal variability across soil types and locations.}

\vspace{0.5cm}

\setstretch{1.2}
\section{Introduction} \label{sec:intro}
Healthy soil plays a critical, yet underappreciated, role in storing and filtering water, sustaining biodiversity, maintaining food production, and mitigating climate change through soil organic carbon sequestration \citep{SoilHealth, SoilReport}. It is estimated that nearly 80\% of carbon in the terrestrial ecosystems of the planet is found in soil \citep{SOC}. Soil water is an important component of soil health, crucial to the supply of water to plants, and is therefore fundamental to agricultural production. It is also a key component in the hydrological cycle, regulating the recharge of groundwater and the flow of water to surface water bodies - both are critical for ecosystem function and human health \citep{SoilDrydown}. In addition, soil moisture is intricately connected to large-scale climate models \citep{SoilDrydown, SDseason}. For example, the soil moisture observations and the identified drydowns were used to evaluate different processes as well as calibrate the associated parameters in the ORCHIDEE (\url{https://orchidee.ipsl.fr/}) land surface model \citep{LSMandSM}.

\subsection{Soil moisture drydown analysis}
The last few decades have seen a growth in research on soil moisture dynamics \citep{SMspatiotemporal}. Soil drydown modelling is one area that has drawn the attention of scientists as more data from underground sensors and satellites have become available. According to \cite{SoilDrydown}, the loss terms in the land water budget (drainage, runoff, and evapotranspiration) are encoded in the shape of the soil moisture drydown curve: the soil moisture time series directly following a precipitation event, during which the infiltration input of water at the soil surface is zero. The soil moisture drydown curve, which is identified from a soil moisture time series, is usually modeled as an exponential decay process \citep{SoilDrydown, SDdiff, SDseason},
\begin{equation}
\theta (t) = \Delta \theta \exp \left(-\frac{t}{\omega} \right) + \theta_{f}
\label{eqn:drydown}
\end{equation}
where $\theta_{f}$ is the estimated lower bound of the soil moisture observations \citep{SoilDrydown, SDcompare}, and $\omega$ reflects the exponential drying rate of the soil and is sometimes referred to as the `temporal e-folding decay' of soil moisture. The temporal variation of the e-folding decay parameter $\omega$ is one of the aspects that soil scientists are interested in. For example, \cite{SDseason} and \cite{SDseason2} investigated the seasonal dynamics and the spatial patterns in the e-folding decay in southeastern South America, leading to a discussion on the importance of effective sampling frequency.

Typically soil drydown modelling requires the soil moisture time series to be manually separated into segments representing the drydown process, i.e., the drydown curves, and then an exponential decay model is fitted to them \citep{SoilDrydown, SDseason}. The results are often snapshot views of the drydown property, characterized by the temporal e-folding decay parameter, or in an interseasonal study, a set of snapshots reflecting the temporal variations of different seasons \citep{SDseason, SDglobal}. Advancements in sensor technology allow scientists to obtain higher frequency time series of soil moisture over longer periods at lower costs, meaning it is now possible to monitor the changes in soil properties through time. Ecological observatories, such as NEON (National Ecological Observatory Network, \url{https://data.neonscience.org/}), are collecting large volumes of such data offering the potential to ask questions about how soil properties, such as those associated with drydown curves, change dynamically. The large volumes of data and continuous monitoring now available present challenges to conventional modelling approaches that rely on the manual extraction of soil dynamics.

Motivated by the current practice of data segmentation, this paper proposes a novel changepoint-based approach to automatically identify the drydown patterns in the soil drying process. In a nutshell, changes caused by sudden increases in soil moisture over a long time series are captured automatically, and the parameters characterizing the drying processes following the sudden increases are estimated simultaneously. Specifically, each segment following a changepoint is modeled using an exponential decay model similar to model (\ref{eqn:drydown}) with segment-specific parameters. This allows the model to capture the temporal variations in the drying process and complements conventional soil drydown modelling. It requires little data pre-processing and can be applied to a soil moisture time series directly, which is attractive when working with large data sets. To identify changepoints, we extend the penalised exact linear time (PELT) method \citep{PELT} to estimate the soil moisture model parameters simultaneously. 

Similar patterns to soil moisture time series (Figure \ref{fig:data-example}) are seen in calcium imaging data in neuroscience. A key problem in computational neuroscience is the inference of the exact times the neuron spiked based on the noisy calcium fluorescence trace. \cite{CaSpike} treat this as a changepoint problem and improvements to their initial approach are made in \cite{CaSpike2}. There are major differences between the neuroscience and soil moisture problems. Whereas neuroscientists are interested in the timing of the spikes, soil scientists are interested in the characteristics of the exponential drying process. Thus whilst the decay parameter is a nuisance in calcium imaging and is integrated out in \cite{CaSpike2}, as their method works with a single parameter (the jump size), the decay is a key parameter characterizing the drying of the soil. Furthermore, the decay parameter in soil may display temporal dynamics as suggested in \cite{SDseason, SDglobal} so there is interest in monitoring it across segments. Alongside this, within the soil moisture drydown models, we may wish to include covariate information e.g., precipitation, temperature, or vegetation, that may allow us to improve the segmentation and better model the individual segments.

\subsection{NEON data}
\begin{figure}[!htb]
\begin{center}
\includegraphics[width=5.5in]{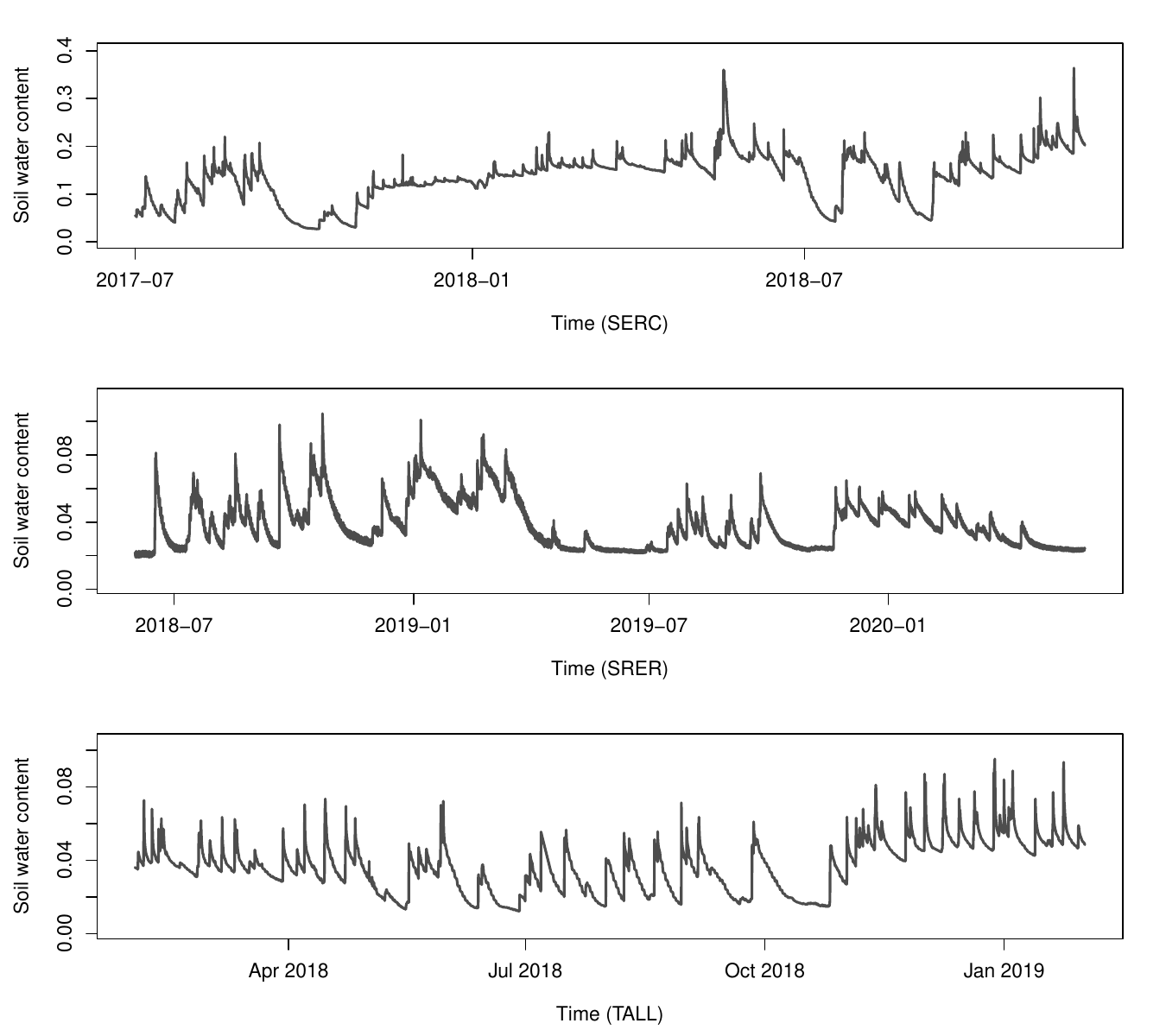}
\captionsetup{labelfont=bf, font=small}
\caption{The hourly soil volumetric water content time series recorded at field site SERC, SRER and TALL from the NEON data portal.}
\label{fig:data-example}
\end{center}
\end{figure}

The development of the changepoint-based modelling approach is motivated by the soil moisture time series from NEON soil water and salinity data product \citep{NEON}. NEON, or National Ecological Observatory Network, is a continental-scale observation facility designed to collect long-term open-access ecological data to better understand how U.S.\ ecosystems are changing. The soil moisture data were collected using the Sentek TriSCAN sensors and available as 1-minute and 30-minute interval time series products \citep{NEONatbd}. Figure \ref{fig:data-example} shows some examples of the hourly (sub-sampled) soil moisture time series recorded at different field sites, including the time series from June 2018 to June 2020 at Santa Rita Experimental Range (SRER) in Arizona, the time series from July 2017 to November 2018 at Smithsonian Environmental Research Center (SERC) in Maryland and the time series from February 2018 to January 2019 at Talladega National Forest (TALL) in Alabama. All these time series display sudden increases in soil moisture followed by drying processes at potentially different decay speeds over time. modelling the drydown characteristics at the three sites has the potential to highlight differences between the soils and their response to different climates and vegetation. Manually identifying and extracting the drydown curves would be very laborious and difficult to deploy rapidly to further sites. The changepoint-based approach we propose provides a solution. The estimated changepoints and parameters provide a dynamic summary of the data, as opposed to the static view from conventional drydown analyses. We hypothesize that, with a sufficiently long time series, the method has the potential to identify long-term changes in the drydown parameters, which could be essential to the study of changing soil health.

The remainder of the paper is divided into five sections. Section \ref{sec:method} introduces the proposed changepoint-based method to model soil moisture time series and the algorithm to implement the method. Section \ref{sec:simulation} presents the simulation study to assess the performance of the method. Section \ref{sec:remarks} discusses the practical aspects of the method, including the use of covariates and penalty learning. Section \ref{sec:application} applies the method to the soil moisture time series from the NEON data portal. Section \ref{sec:discussion} concludes the paper and discusses some directions for future work.

\section{The changepoint method for soil moisture time series} \label{sec:method}
\subsection{The proposed model for soil moisture dynamics} \label{sec:newmodel}
Denote the observed soil moisture at time $t$ by $Y_{t}$. Denote the set of time points where sudden increases in soil moisture occur as $\tau_{i}$, $i=1, \cdots, k$. The dynamics in soil moisture can be described using 
\begin{align}
\label{eqn:smSpike}
Y_{t} & = X_{t} + \epsilon_{t} \, , \;\; \epsilon_{t} \sim \mathrm{i.i.d} \; \mathcal{N} (0, \sigma^{2})\\
X_{t} & = \phi_{t} \, X_{t-1} + \Delta_{t} \; .  \nonumber
\end{align}
Here $X_{t}$ is the underlying soil moisture, $\phi_{t}$ ($0 < \phi_{t} < 1$) is the soil moisture decay parameter, and $\Delta_{t}$ is the increase in soil moisture, which takes a positive value at time points $\tau_{i}$, $i=1, \cdots, k$ and 0 otherwise. The second equation in model (\ref{eqn:smSpike}) translates to an exponential decay model for the segment between two sudden increases. That is, for $t \in (\tau_{i}+1, \tau_{i+1})$, 
\begin{equation}
Y_{t} = \alpha_{0i} + \alpha_{1i} \phi_{i}^{(t-\tau_{i})}  + \epsilon_{t} \, . 
\label{eqn:expDecay0}
\end{equation}
Here $\alpha_{0i}$ ($\alpha_{0i} > 0$) is the asymptotic soil moisture and $\alpha_{0i} + \alpha_{1i}$ ($\alpha_{1i} > 0$) is the soil moisture after $\tau_{i}$. Unlike the decay of calcium concentration, soil moisture decreases at different speeds over different periods. This is a result of the temporal variation in the elements that affect the speed at which the soil loses water, e.g.\ temperature and vegetation. It carries interesting information on soil moisture dynamics. To reflect this feature, a segment specific decay parameter $\phi_{i}$ is used in the model, i.e. $\phi_{t} = \phi_{i}$ for $t \in (\tau_{i}+1, \tau_{i+1})$, $i=1, \cdots, k$. Depending on the properties of the soil, the asymptotic soil moisture can be fixed throughout a time series or be segment-specific.

To make the fitting of the exponential decay model (\ref{eqn:expDecay0}) easier, a re-parameterisation $\phi_{i} = \exp \{ -\exp(\gamma_{i}) \}$ is used, giving
\begin{equation}
Y_{t} = \alpha_{0i} + \alpha_{1i} \exp \{ - \exp(\gamma_{i}) \, (t-\tau_{i}) \}  + \epsilon_{t} \, . 
\label{eqn:expDecay}
\end{equation}
This removes the constraint on $\phi_{i}$ so that $\exp \{ - \exp(\gamma_{i}) \} \in (0, 1)$ for all $\gamma_{i} \in \mathbb{R}$. Note that $\gamma_{i}$ is essentially a reparameterisation of the e-folding decay parameter $\omega$ in the soil drydown model (\ref{eqn:drydown}). In other words, $1/\exp(\gamma)$ is equivalent to $\omega$ if $t$ in model (\ref{eqn:drydown}) and (\ref{eqn:expDecay}) have the same unit. 

The parameters of the exponential decay model (\ref{eqn:expDecay}) can be estimated by minimising the non-linear least square (NLS) fit. Iterative algorithms, such as Gauss-Newton, Newton-Raphson, and Levenberg-Marquardt can be used to solve the optimisation problem \citep{nlsBook}. These algorithms can be implemented using various R \citep{R} functions, e.g.\ the \texttt{nls} function which implements the Guass-Newton algorithm and the \texttt{port} algorithm \citep{NL2SOL}, or \texttt{nlfb} function from package \texttt{nlmrt} \citep{nlmrt} which uses the Nash variant of the Levenberg-Marquardt algorithm \citep{NLMoptim}. 

The negative log-likelihood of the estimated model (\ref{eqn:expDecay}) is used as the cost function of the changepoint detection problem. Note that the cost function is a function of multiple parameters, $\alpha_{0i}, \alpha_{1i}$ and $\gamma_{i}$. As a result, functional pruning, which was developed for uni-parameter cost functions, is not an appropriate choice \citep{DPA, FPOP}. The additional effort required to identify the multi-dimensional region where the multivariate cost function attains its minimum at each step undermines the computational efficiency of functional pruning. Therefore, this paper chooses to develop a changepoint detection procedure based on the penalised exact linear time (PELT) method \citep{PELT}. The PELT method is flexible and computationally efficient, making it suitable for the soil moisture time series, which typically consists of 10,000 to 20,000 time points. Details of the PELT method and its applicability to the problem in this paper are described in section \ref{sec:algorithm}. 

In addition, lower and upper limits may be applied to the asymptotic soil moisture parameter $\alpha_{0i}$ and the increase parameter $\alpha_{1i}$, to ensure valid soil moisture values and a positive increase. The positive constraint on the increase of soil moisture translates to $\alpha_{0i} + \alpha_{1i} > \alpha_{0i+1}$. The constraints on parameters are treated as lower and upper bounds in NLS optimisation.

\subsection{Model estimation using PELT} \label{sec:algorithm}
The optimisation goal is to identify the set of changepoints, $0=\tau_{0} < \tau_{1} < \cdots < \tau_{k} < \tau_{k+1} = n$, that minimise the overall penalised cost function
\begin{equation}
\sum_{i=0}^{k} \mathcal{C} (Y_{(\tau_{i}+1) : \tau_{i+1}}) + \lambda f(k) \; ,
\label{eqn:smSpike-cpt}
\end{equation}
where the cost function $\mathcal{C} (Y_{(\tau_{i}+1) : \tau_{i+1}})$ is
\begin{equation*}
(\tau_{i+1} - \tau_{i}) \left\{ \log(2 \pi) + 1 + \sum_{t=\tau_{i}+1}^{\tau_{i+1}} \left\{ Y_{t} - \hat{\alpha}_{0i} - \hat{\alpha}_{1i} \exp ( - \exp(\hat{\gamma}_{i}) \; (t-\tau_{i}) ) \right\}^{2} \right\} \; ,
\label{eqn:smSpike-cost}
\end{equation*}
which is twice the negative log-likelihood of the exponential decay model (\ref{eqn:expDecay}) fitted to the segment $Y_{(\tau_{i}+1)}, \cdots, Y_{\tau_{i+1}}$, and the penalty function is $f(k) = k$, the number of changepoints. Other types of penalties are available, e.g.\ the modified BIC \citep{mbic}. 

The PELT method by \cite{PELT} starts with the recursive computation of the overall cost function of the data up to time point $s$, $s=1, \cdots, n$. Denoted this cost function as $F(s)$, the recursion is
\begin{equation*}
F(s) = \displaystyle{\min_{\bm{\tau} \in \mathcal{T}_{s}}} \left\{ \sum_{i=0}^{m} \mathcal{C} (Y_{(\tau_{i}+1) : \tau_{i+1}}) + \lambda k \right\} = \displaystyle{\min_{0 \leq \tau < s}} \left\{ F(\tau) + \mathcal{C} (Y_{(\tau+1) : s}) + \lambda \right\} \; ,
\end{equation*}
where $\bm{\tau} = (\tau_{1}, \cdots, \tau_{m})$, $\mathcal{T}_{s}$ is the set of $\{ \bm{\tau}: 0=\tau_{0} < \tau_{1} < \cdots < \tau_{m} < \tau_{m+1} = s \}$, and $\tau$ is the last changepoint before $s$. Instead of searching through all candidate time points $0 \leq \tau < s$ for the optimal solution to $\tau$, the algorithm prunes the candidate time points that can never be the last optimal changepoint for data $Y_{1:s}$, and searches only within a reduced set of candidate time points. Specifically, the pruning criterion \citep{PELT} is, for all $t < t' < s$ satisfying
\begin{equation}
\mathcal{C} (Y_{(t+1) : t'}) + \mathcal{C} (Y_{(t'+1) : s}) + K \leq \mathcal{C} (Y_{(t+1) : s})
\label{eqn:PELT1}
\end{equation}
for some constant $K$, the time point $t$ can never be the last optimal changepoint prior to time point $s$ if
\begin{equation}
F(t) + \mathcal{C} (Y_{(t+1) : t'}) + K \geq F(t') \, .
\label{eqn:PELT2}
\end{equation}
Consequently, all time points $t$ that satisfy condition (\ref{eqn:PELT2}) can be removed from the search and the computational cost is reduced.

Under the i.i.d.\ Normal distribution assumption of $\epsilon_{t}$ in the exponential decay model (\ref{eqn:expDecay}), twice the negative log-likelihood of the model satisfies the inequality (\ref{eqn:PELT1}) with $K=0$ \citep{PELT}. Intuitively, consider adding a changepoint at $t'$ between $t$ and $s$ while keeping the parameters in the exponential decay model unchanged. This gives the equivalence condition of (\ref{eqn:PELT1}) with $K=0$. Any updated parameters that reduce either $\mathcal{C} (Y_{(t+1) : t'})$ or $\mathcal{C} (Y_{(t'+1) : s})$ will reduce the overall cost. Hence, a changepoint detection procedure based on PELT can be established, which is depicted in Algorithm \ref{alg:PELT1}.

\begin{algorithm}[!htb]
\SetKwInOut{Input}{Input}
\SetKwInOut{Output}{Output}

\Input{
data $Y_{1}, \cdots, Y_{n}$; \\
cost function $\mathcal{C} (Y_{t:s})$; \\
minimum segment length $l \geq 3$, depending on the application; \\
penalty parameter $\lambda$ that penalises the number of changepoints; \\
constant $K$ that satisfies the inequality (\ref{eqn:PELT2}); \\
}

\BlankLine
\textbf{Initialise} \;
\For{$t = 1, \cdots, 2l-1$} {
fit the exponential decay model to $Y_{1:t}$ and compute the cost function $F(t) = \mathcal{C} (Y_{1:t}) + \lambda$\;
initialise the set of changepoint $\mathrm{cpt}(t) = \{ 0 \}$ \;
initialise the set of candidate changepoints $\mathrm{R}_{t} = \{ 0 \}$ \;
}
set $F(0) = -\lambda$ and $\mathrm{R}_{2l} = \{ 0, l \}$ \;

\BlankLine
\textbf{Iterate} \;
\For{$t \leftarrow 2l$ \KwTo $n$}{
1. fit the exponential decay model to $Y_{(\tau+1):t}$, $\tau \in \mathrm{R}_{t}$, and compute $F(t) = \min_{\tau \in \mathrm{R}_{t}} \left\{ F(\tau) + \mathcal{C} (Y_{(\tau+1):t}) + \lambda \right\} $ \;

2. find $\tau^{\ast} = \mathrm{arg} \min_{\tau \in \mathrm{R}_{t}} \left\{ F(\tau) + \mathcal{C} (Y_{(\tau+1):t}) + \lambda \right\}$ \;

3. update $\mathrm{cpt}(t) = \{ \mathrm{cpt}(\tau^{\ast}), \tau^{\ast} \}$ \;

4. update $\mathrm{R}_{t+1} = \left\{\{ t-l+1 \} \cup \{\tau \in \mathrm{R}_{t}: F(\tau) + \mathcal{C} (Y_{(\tau+1):t}) + K \leq F(t) \} \cup \mathrm{S}_{t} \right\}$, where $\mathrm{S}_{t} =  \left\{ \tau \in \mathrm{R}_{s}, s = t, \cdots, t-l+2 : F(\tau) + \mathcal{C} (Y_{(\tau+1):s}) + K > F(s) \right\} $ \;
}

\BlankLine
\Output{
a set of changepoints $\mathrm{cpt}(n) = \{ 0, \tau_{1}, \cdots, \tau_{k}, n \}$; \\
estimated exponential decay model parameters $\hat{\alpha}_{0i}, \hat{\alpha}_{1i}, \hat{\gamma}_{i}$, $i = 1, \cdots, k+1$. \\
}

\caption{PELT for soil moisture time series} \label{alg:PELT1}
\end{algorithm}

Sometimes, the non-linear least square estimation of the model (\ref{eqn:expDecay}) does not produce a converged result. In these situations, the cost function of the segment is set to a very large value to represent an infinite cost, and the PELT iteration is modified slightly. During the iteration, both $F(\tau)$ and $\mathcal{C} (Y_{(\tau+1):t})$ could be infinite for a candidate changepoint $\tau$. When the `historical' cost $F(\tau)$ is infinite, then $\tau$ can never be the last optimal changepoint prior to $t$ and hence it should be pruned. On the contrary, when $F(\tau)$ is finite, but $\mathcal{C} (Y_{(\tau+1):t})$ is infinite, there is a possibility that the model fitted to the segment starting from $\tau+1$ will converge when more observations are added to the segment. That is, $\mathcal{C} (Y_{(\tau+1):(t+\delta)})$ may be finite for some $\delta \geq 1$. Therefore, no pruning is applied to those $\tau$ values and they are all kept for the next iteration. The modified iteration is given in the supplemental document.

\section{Simulation study} \label{sec:simulation}
To investigate the performance of the method developed above, a simulation study is carried out. Two problems of particular interest in this case are, (1) whether the algorithm can identify the locations of the sudden increases in the time series in different scenarios, and (2) whether the method can produce a reasonable estimation of the model parameters, in particular $\gamma$.

\subsection{Simulation design} \label{sec:simDesign}
The observed soil moisture time series sometimes display temporal patterns in the frequency of the sudden increases, e.g.\ more frequent increases during the rainy summer season than the dry winter period. The sudden increases in soil moisture may also appear at very different scales. For example, there can be a series of smaller increases during a long large-scale drying process as in the top penal of Figure \ref{fig:data-example}. A few more examples of different temporal patterns are given in the supplemental document. Based on these features, three scenarios are considered in terms of the frequency of the sudden increases, (a) randomly distributed over time, (b) following a temporal pattern where one part of the time series has more frequent increases than the rest, (c) large scale sudden increases randomly distributed over time, along with small scale increases over a long drying period. Each of the scenarios will be paired with two noise levels, giving six scenarios in total. 

There may be temporal variation in the drying rate in a long time series, which can be attributed to e.g., seasons, vegetation, and human activities. This is reflected in the simulated time series by alternating slow-drying and fast-drying periods. Despite the interest in the decay parameter, this type of variation does not affect the changepoint detection procedure. Therefore, the temporal variation of the decay parameter is fixed across all scenarios. 

\begin{table}[!htbp]
\begin{center}
\captionsetup{labelfont=bf, font=small}
\caption[Simulation scenarios]{The specifications of the simulation scenarios.}
\small
\begin{tabular}{p{2.1cm}|cccc}
\toprule[1.5pt]
  & \textbf{Spikes} & \textbf{Drying rate} & \textbf{Noises} & \textbf{Replicates} \\
\hline
 \textbf{Scenario 1a} & 1 Poisson process & 2 uniform distributions & $\sigma = 0.0005$ & 200 \\
 \textbf{Scenario 2a} & 2 Poisson process & 2 uniform distributions & $\sigma = 0.0005$ & 200 \\
 \textbf{Scenario 3a} & 2 Resolutions (Poisson) & 2 resolutions (uniform) & $\sigma = 0.0005$ & $100 \times 2$ penalties \\
\midrule
 \textbf{Scenario 1b} & 1 Poisson process & 2 uniform distributions & $\sigma = 0.001$ & 200 \\
 \textbf{Scenario 2b} & 2 Poisson process & 2 uniform distributions & $\sigma = 0.001$ & 200 \\
 \textbf{Scenario 3b} & 2 Resolutions (Poisson) & 2 resolutions (uniform) & $\sigma = 0.001$ & $100 \times 2$ penalties \\
\bottomrule[1.5pt]
\end{tabular}\label{tab:scenario}
\end{center}
\end{table}

Time series of length 5000 are generated using the following steps. First, the changepoints are simulated from different Poisson distributions. In particular, changepoints in scenarios 1a and 1b are generated from a single Poisson distribution. Changepoints in scenarios 2a and 2b are simulated from two Poisson distributions with different intensity parameters to reflect the temporal patterns. Changepoints in scenarios 3a and 3b are simulated by two nested Poisson processes, one over the entire time span $T=5000$ with lower intensity, and the other over a long decaying period with higher intensity. Then the drying rates, the spikes and the asymptotic parameters are simulated from various uniform distributions. Finally, Gaussian random noises are added to the time series. A summary of the specifications of the six scenarios is given in Table \ref{tab:scenario}. Details of the simulation procedure and an example of the simulated time series from each of the six scenarios are given in the supplemental document. 

The simulation was implemented in R using the code developed by the authors. The following statistics were computed to investigate the performance of the method, (a) the true positive (TP) rate and false positive (FP) rate of the changepoint detection, (b) the distance between the set of estimated changepoints and the set of true changepoints, (c) the difference between the fitted time series and the true simulated time series quantified as the root mean squared errors (RMSE), (d) the difference between the estimated drying rates and the true simulated drying rates quantified as the RMSE.

\subsection{Summary of simulation results} \label{sec:simResult}

Table \ref{tab:TPFPrate} shows the averaged true positive rates and false positive rates overall simulation replicates for six simulation scenarios. The true positive rates reached over 90\% for scenarios 1a, 1b, 3a large and 3b large. The true positive rates are relatively lower (82.39\% and 85.96\% respectively) in scenarios 3a small and 3b small. This is due to the challenges in estimating smaller-scale changes as the signals are much weaker. The results improved when a relaxed version of true positive is considered, i.e.\ an estimated changepoint $\hat{\tau}_{j}$ has a match with a true changepoint $\tau_{i}$ if $\hat{\tau}_{j} \in (\tau_{i}-10, \tau_{i}+10)$. Increasing noise standard errors did not appear to affect the changepoint detection, which may be explained by the fact that the gaps between small and large noise levels are not distinctive enough to cause major differences. The false positive rates are low across all scenarios with the majority of replicates smaller than 0.1\%, regardless of counting the exact match or a match within the $(\tau_{i}-10, \tau_{i}+10)$ intervals. Over 1/3 of the replicates in scenarios 1a, 2a, 1b, and 2b have false positive rates of 0. This is slightly lower in scenarios 3a and 3b when the small-scale increases are considered. 

\begin{table}[!htbp]
\begin{center}
\captionsetup{labelfont=bf, font=small}
\caption{Average true positive rates and false positive rates (in \%), and those calculated under the relaxed matching condition (i.e.\ within $\pm 10$ time points of the true changepoint)}
\small
\begin{tabular}{p{2cm}|p{1.5cm}p{1.5cm}|p{2cm}p{2cm}}
\toprule[1.5pt]
  & \textbf{TP} & \textbf{FP} & \textbf{TP ($\pm10$)} & \textbf{FP ($\pm10$)}  \\
\hline
 \textbf{S1a} & 91.96\% & 0.02\% & 94.40\% & 0.01\% \\ 
 \textbf{S2a} & 89.71\% & 0.02\% & 92.36\% & 0.01\% \\ 
 \textbf{S3a (small)} & 82.39\% & 0.05\% & 86.12\% & 0.03\% \\ 
 \textbf{S3a (large)} & 95.76\% & 0.04\% & 96.45\% & 0.04\% \\ 
\midrule
 \textbf{S1b}  & 92.05\% & 0.02\% & 94.77\% & 0.01\% \\
 \textbf{S2b}  & 89.71\% & 0.02\% & 92.51\% & 0.01\% \\ 
 \textbf{S3b (small)} & 85.96\% & 0.04\% & 87.89\% & 0.02\% \\ 
 \textbf{S3b (large)} & 95.91\% & 0.02\% & 96.35\% & 0.02\% \\
\bottomrule[1.5pt]
\end{tabular}\label{tab:TPFPrate}
\end{center}
\end{table}

For a more comprehensive comparison of the estimated and true changepoints, a distance developed by \cite{DistCpt} was computed to investigate the dissimilarity between the two configurations, e.g.\ the true changepoints $\bm{\tau} = \{\tau_{1}, \cdots, \tau_{m} \}$ and the estimated changepoints $\bm{\eta} = \{\eta_{1}, \cdots, \eta_{k} \}$. It is defined as
\begin{equation*}
d(\bm{\tau}, \bm{\eta}) = | m - k| + \min \{ \mathcal{A}(\bm{\tau}, \bm{\eta}) \} \; ,
\label{eqn:distCpt}
\end{equation*}
where $m$ and $k$ are the number of changepoints in each set, and
\begin{equation*}
\mathcal{A}(\bm{\tau}, \bm{\eta}) = \sum_{i=1}^{m} \sum_{j=1}^{k} c_{ij} I_{ij} = \sum_{i=1}^{m} \sum_{j=1}^{k} \frac{( \tau_{i} - \eta_{j} )}{N} I_{ij}, \; 
\end{equation*}
is the overall cost of assigning $\eta_{j}$ to $\tau_{i}$, $j=1, \cdots, k$, $i=1, \cdots, m$. To be specific, $I_{ij} = 1$ if $\eta_{j}$ is assigned to $\tau_{i}$ and $I_{ij} = 0$ otherwise, following a linear assignment problem. Note that when $m \neq k$, not all $\tau_{i}$ and $\eta_{j}$ are paired; when there is a perfect match between the two sets of changepoints, $d(\bm{\tau}, \bm{\eta}) = 0$. Such a distance accounts for the dissimilarity in both the number and the locations of the true and estimated changepoints, thus providing a more comprehensive quantification of the differences. The distances are presented in Table \ref{tab:dist}. It appears that most of the scenarios have relatively small distances, apart from scenario 3b where some smaller-scale changepoints are missed when the noise level is higher.

The RMSE of the fitted time series is computed as a measure of the overall fit of the estimated model. Alongside this, the RMSE of the estimated decay parameter $\gamma$ is also computed to investigate how the method retrieved the key parameter in the exponential decay model. The results are shown in Table \ref{tab:dist}. The overall fit of the model was reasonable for all scenarios. The RMSE of the PELT runs using a larger penalty in scenarios 3a and 3b are among the highest, which is expected as they were designed to capture only the large-scale increases. It also seems difficult to retrieve the decay parameters $\gamma$ from the small-scale increases in scenarios 3a and 3b, which again is as expected. The RMSEs of $\gamma$ are an order of magnitude smaller in the rest of the scenarios than the two most challenging ones. 

\begin{table}[!htbp]
\begin{center}
\captionsetup{labelfont=bf, font=small}
\caption{Summary statistics of the distance measure between 2 sets of changepoints, the root mean squared errors (RMSE) of the fitted time series and the RMSE of the estimated decay parameter $\gamma$.}
\small
\begin{tabular}{p{2cm}|ccc|ccc|ccc}
\toprule[1.5pt]
  & & \textbf{Distance} & & & \textbf{RMSE}  & & & \textbf{RMSE $\gamma$} & \\
  & 10\% & median & 90\% & 10\% & median & 90\% & 10\% & median & 90\%\\ 
\hline
 \textbf{S1a} & 0 & 0.0015 & 1.2325 & 0.0005 & 0.0023 & 0.0068 & 0.0001 & 0.0016 & 0.0936 \\ 
 \textbf{S2a} & 0 & 0.0227 & 2.1863 & 0.0005 & 0.0022 & 0.0053 & 0.0002 & 0.0124 & 0.0926 \\ 
 \textbf{S3a (small)} & 0.0053 & 1.1093 & 5.0979 & 0.0006 & 0.0009 & 0.0047 & 0.0147 & 0.0844 & 0.1341 \\ 
 \textbf{S3a (large)} & 0 & 1.0019 & 4.0000  & 0.0019 & 0.0033 & 0.0055 & 0.0006 & 0.0063 & 0.0401 \\ 
\midrule
 \textbf{S1b}  & 0 & 0.0053 & 1.2023 & 0.0009 & 0.0024 & 0.0073 & 0.0002 & 0.0023 & 0.0980 \\
 \textbf{S2b}  & 0 & 0.0343 & 2.1302 & 0.0009 & 0.0024 & 0.0054 & 0.0003 & 0.0147 & 0.0934 \\ 
 \textbf{S3b (small)} & 0.0128 & 3.0703 & 23.0000 & 0.0010 & 0.0013 & 0.0049 & 0.0110 & 0.0713 & 0.1379 \\ 
 \textbf{S3b (large)} & 0 & 1.0000 & 2.0912 &  0.0024 & 0.0036 & 0.0062 & 0.0006 & 0.0032 & 0.0174\\
\bottomrule[1.5pt]
\end{tabular}\label{tab:dist}
\end{center}
\end{table}

\begin{figure}[!htb]
\begin{center}
\includegraphics[width=5.8in]{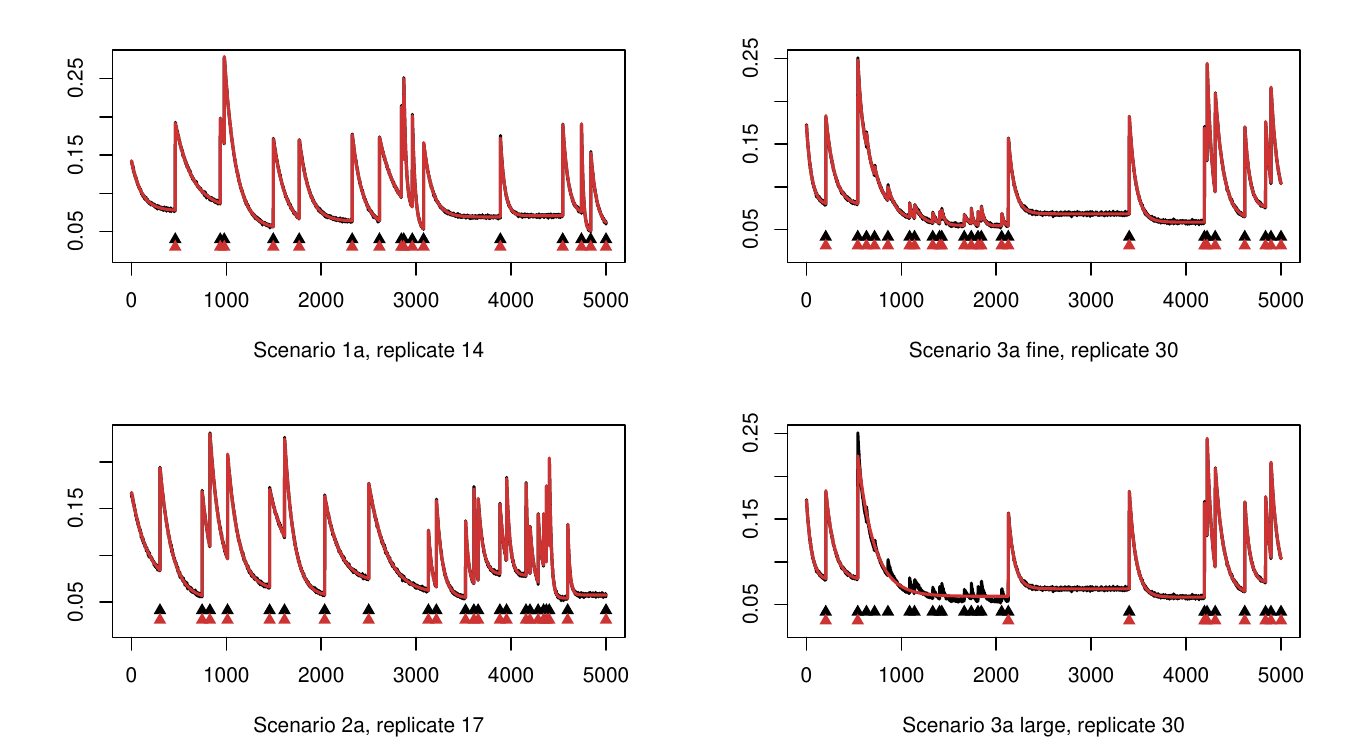}
\captionsetup{labelfont=bf, font=small}
\caption{Examples of simulation replicates with high true positive rate from scenarios 1a, 2a and 3a. The black curve and the black triangles represent the simulated time series and the true changepoints. The red curve and the red triangles represent the fitted time series and the estimated changepoints.}
\label{fig:sim-good}
\end{center}
\end{figure}

\begin{figure}[!htb]
\begin{center}
\includegraphics[width=5.8in]{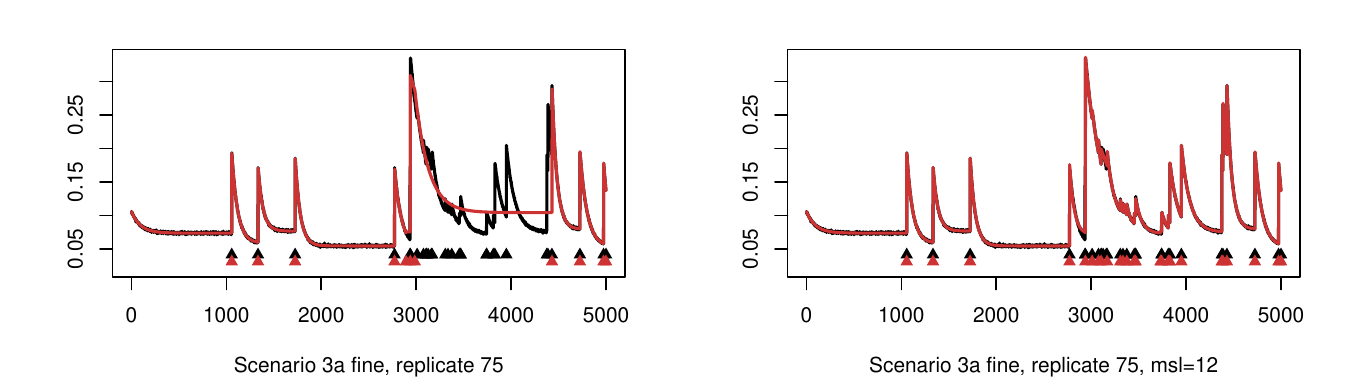}
\captionsetup{labelfont=bf, font=small}
\caption{An example of a simulation replicate with low true positive rate (left panel), and the improved result using smaller minimum segment length (right panel) from scenario 3a. The black curve and the black triangles represent the simulated time series and the true changepoints. The red curve and the red triangles represent the fitted time series and the estimated changepoints.}
\label{fig:sim-bad}
\end{center}
\end{figure}

Finally, a few examples of the replicates from different scenarios are presented to give intuition to the averages presented in the tables. Examples of replicates with high true positive rates and small mean squared errors from scenarios 1a, 2a and 3a are given in Figure \ref{fig:sim-good}. They represent the performance of the proposed method in the majority of the replicates. There are a few situations where the method failed to achieve a satisfactory fit, either in terms of the true positive rate or the estimated decay parameter. An example of the replicate with low true positive rates is shown in the left panel of Figure \ref{fig:sim-bad}. Here the lack of fit was the result of the minimum segment length used (24) being larger than the distance between two adjacent increases. Hence, the method missed the changepoints and created a knock-on effect on some later time points. This can be improved by simply reducing the minimum segment length, see for example the right panel of Figure \ref{fig:sim-bad}, where the minimum segment length is 12. The lack of fit due to the difficulty in capturing the smaller scale patterns in scenarios 3a and 3b may be improved by changing the penalty parameter $\lambda$ in the optimisation problem. In the simulation study, both the penalty and the minimum segment length were fixed for all replicates; whereas in real application, these values will be tuned according to the problem.

\section{Practical Considerations} \label{sec:remarks}
The simulation study in section \ref{sec:simulation} demonstrates the ability of the proposed method to detect structural changes and exponential decay parameters in different types of time series of repeated sudden increases and decays. In real applications, however, there could be complications that require careful consideration. Changepoint detection may be improved by taking into account additional information. In this section, several aspects of the proposed method are discussed, including the possibility of adding covariates to the model and the selection of penalty parameters.

Precipitation is considered to be one of the most important drivers of soil moisture increase. Although it is not equivalent to the amount of water infiltrating into the soil due to other pathways of water loss, such as surface run-off \citep{SoilPhysics, SoilDrydown}, it is expected to correlate with the locations and frequencies of the peaks. When relevant precipitation data are available, one could consider including precipitation as a covariate of the exponential decay model. For example, the precipitation time series $X_{t}$ may be used directly to help model the peaks of soil moisture as
\begin{equation*}
  Y_{t} = \alpha_{0i} + (\alpha_{1i} + \beta_{i} X_{t}) \exp \{ - \exp(\gamma_{i}) \, (t-\tau_{i}) \}  + \epsilon_{t} \; .
\end{equation*}
Alternatively, it may be converted into an indicator variable $Z_{t}$ and included in the model as
\begin{equation*}
  Y_{t} = \alpha_{0i} + (\alpha_{1i} + \bm{Z}_{ti}^{\top}\bm{\beta}_{i}) \exp \{ - \exp(\gamma_{i}) \, (t-\tau_{i}) \}  + \epsilon_{t} \; ,
\end{equation*}
where $\bm{Z}_{ti}^{\top} = (Z_{ti}^{1}, \cdots, Z_{ti}^{P})$, with $Z_{ti}^{p} = \mathbbm{1}(t \geq z_{p})$ for $p=1, \cdots, P$ and $z_{1}, \cdots, z_{P}$ being the timings of rainfall instances within segment $i$. This changes the content of the changepoints from the times when soil moisture peaks to the times when the decay rate or the asymptotic level changes, or the time when the moisture level peaks without rainfall. However, they are still suitable for the investigation of the temporal dynamics of soil moisture drydown. For this approach to work, the precipitation time series needs to reflect the rainfall-infiltration pattern relatively accurately, which is challenging for field data. In addition, sometimes the precipitation becomes surface run-off, which does not contribute to the change of soil water content.

The proposed method relies on tuning parameters, such as the penalty parameter and the minimum segment length. The penalty used in sections \ref{sec:method} and \ref{sec:simulation} is the BIC. Other choices are available and equally theoretically valid. In the simulation study, the penalty parameter was fixed within each scenario. In practice, the model does not reflect the full reality of the process and thus the penalty parameter is not optimal and may need to be selected based on the feature of the time series. For a systematic approach, \cite{PenaltyCpt} introduces the CROPS algorithm, which efficiently computes the changepoint problem for a range of penalties. Different types of adaptive penalties have also been developed in the literature, such as \cite{PenaltyInterval} and \cite{LearnPenalty} where the penalty parameter is estimated to match the annotations of changepoints by experts. Such annotations are usually not available for the soil moisture time series. However, one may use climate data to create pseudo annotations and carry out penalty learning. One potential choice is precipitation.

To implement this we could assume that the locations and frequencies of the peaks correlate with the timings and frequencies of {\textit{some}} rainfall events. Then the precipitation time series, after applying a certain threshold or transformation, may be used as the experts' annotations. Comparing the detected numbers and/or locations of changepoints under different penalty parameters to the annotated changepoints can provide some indication of the choice of an appropriate penalty parameter. For example, the penalty learning method in \cite{PenaltyInterval} compares the possible number of changepoints annotated by experts in different regions of the time series to the number of changepoints detected by the algorithm given penalty $\lambda$ in corresponding regions. The loss function to minimise is 
\begin{equation*}
    \mathcal{E}(\lambda) = \sum_{(r, a) \in \{ R, A \}} \mathbbm{1} \left( \left| \bm{\tau}^{(\lambda)} \bigcap r \right| \notin a \right) ,
    \label{eqn:PenLoss}
\end{equation*}
where $\bm{\tau}^{(\lambda)}$ is the optimal set of changepoints given $\lambda$, $R$ is a set of regions on the time axis and $A$ is a set of annotations on the possible number of changepoints within each region in set $R$. The regions do not necessarily cover the entire span of the time series, which means the missing gaps in the precipitation time series are not problematic. Alternatively, the method in \cite{LearnPenalty} selects the appropriate penalty $\lambda$ through minimising the excess penalised risk, 
\begin{equation*}
    \mathcal{E}(\lambda) = \mathcal{R}\left( \lambda; Y_{1:n}, \bm{\tau}^{xpt} \right) - \mathcal{R}\left( \lambda; Y_{1:n}, \bm{\tau}^{(\lambda)} \right) \; ,
    \label{eqn:PenRisk}
\end{equation*}
where $\mathcal{R}(\lambda; Y_{1:n}, \bm{\tau})$ is the risk function (i.e., overall cost) of segmenting the data according to $\bm{\tau}$, and $\bm{\tau}^{xpt}$ is the collection of the changepoints labelled by the experts. The challenge of implementing such a penalty learning method is the creation of the annotations, i.e., converting the precipitation time series into experts' annotations. The conversion may be case-specific and may require additional information about the field site. However, it does not rely on distinctive rainfall-infiltration patterns. It only requires the rainfall instances to correlate with the occurrences of peaks to some extent. Therefore, precipitation data that are not suitable as covariate data may still be fit for annotations. An example of creating annotations and implementing the two penalty learning methods using NEON soil moisture and precipitation data is given in the supplemental document.


The minimum segment length is used in the proposed method to put a realistic lower bound on the soil drying time and to allow enough data in the NLS estimation of the exponential decay model. Some of the changepoints will inevitably be missed if multiple increases occur within the distance of the minimum segment length, as noted in the simulation study above and also in experiments using real data. In such situations, the estimation of the decay parameter could be problematic. It is possible to reduce the minimum segment length, but whether this is appropriate depends on the properties of the soils and the minimum number of observations needed to adequately fit the exponential decay model. In addition, multiple sudden increases within a very short window may be associated with sensor noise so it is not always appropriate to try to recover these.

\section{Analysing NEON soil moisture time series} \label{sec:application}

In this section, the changepoint model was applied to the soil moisture time series from the NEON data portal. Soil water and salinity data have been collected in 46 field sites across the U.S.. Here three terrestrial field sites with contrasting features: the Smithsonian Environmental Research Center (SERC) in Maryland, the Santa Rita Experimental Range (SRER) in Arizona, and the Talladega National Forest (TALL) in Alabama, are investigated.

A full description of the three field sites can be found at \url{https://www.neonscience.org/field-sites}, but their characteristics are summarised here for the readers' convenience. The Smithsonian Environmental Research Center is located in, Maryland on the Rhode River, a sub-estuary of the Chesapeake Bay. The climate is temperate and humid, with an average annual temperature of $13.6^{\circ}\mathrm{C}$ ($56.5^{\circ}\mathrm{F}$) and a mean annual precipitation of 1075mm. Soils are formed into fluvial marine deposits with some areas of overlying alluvium and loess and the vegetation is dominated by coastal hardwood forests and cropland. The Santa Rita Experimental Range, located in the Sonoran Desert, Arizona, is characterized by a semi-arid, hot climate. The mean annual temperature is $19.3^{\circ}\mathrm{C}$ ($67^{\circ}\mathrm{F}$). The Sonoran Desert is wetter than most deserts with a mean annual precipitation of 346.2mm each year which is distributed in two wet periods. Diurnal temperature swings of up to $32^{\circ}\mathrm{C}$ ($89.6^{\circ}\mathrm{F}$) are common. The soils found at SRER are those typical of desert regions - they are mostly composed of alluvial deposits from the Santa Rita Mountains. Vegetation at the site is dominated by drought-resistant, thorny species. The Talladega National Forest is located in west-central Alabama. It has a subtropical climate with hot summers, mild winters, and year-round precipitation. This warm, moist air contributes to the formation of convection storms and thunderstorms in the region, causing major precipitation pulses and flooding. The area is subject to tornadoes and hurricanes. The average annual temperature is $17.2^{\circ}\mathrm{C}$ ($62.9^{\circ}\mathrm{F}$) and the average annual precipitation is about 1380mm. The soils in TALL are primarily sand, clay, and mudstone formed from undifferentiated marine segments. The vegetation at TALL is dominated by conifers, with some areas of intermixed conifers, hardwoods, bottomland hardwoods, and wetlands.  

Soil moisture measurements are made in vertical profiles consisting of up to eight depths in five instrumented soil plots at each site. The data are presented as 1-minute and 30-minute averages. Here the 30-minute data product is used and the data are further sub-sampled to 1-hour time series for the changepoint analysis\footnote{In conventional soil drydown modelling, the temporal resolutions of the data are usually lower. For example, analysis using remote sensing data would be using daily data or data of coarser resolution.}. 

The location with the fewest missing observations was selected from each field site, and the period with no large missing gap was selected. These are, 1 July 2017 to 30 November 2018 at location 1 for field site SERC, 1 June 2018 to 31 May 2020 at location 4 for field site SRER, and 1 February 2018 to 31 January 2019 at location 5 for field site TALL (see Figure \ref{fig:data-example}). Linear interpolation was applied to fill the small amounts of missing data within each time series.

To begin with, the constraints on model parameters are set and the values of the tuning parameters are selected. (a) An upper cap of 0.4 was applied to the soil moisture time series from the NEON data portal. As a result, the same upper bound was introduced to the parameters $\alpha_{0i}$ and $\alpha_{1i}$ in model (\ref{eqn:expDecay}) during the optimisation of the non-linear least squares. (b) The minimum segment length was chosen to be 24 hours. (c) It was required that the size of the sudden increases be greater than 0.001. This value was used to filter out the sensor noise, which is on the scale of $10^{-4}$ \citep{NEONatbd}. (d) Considering jointly the residuals of the fitted time series and the experimental result from penalty learning using the precipitation time series, it appeared that a penalty around 250 would be appropriate for the soil moisture time series from site SERC, a penalty around 200 for site SRER, and a penalty around 200 for site TALL.  

\begin{figure}[!htb]
\begin{center}
\includegraphics[width=6in]{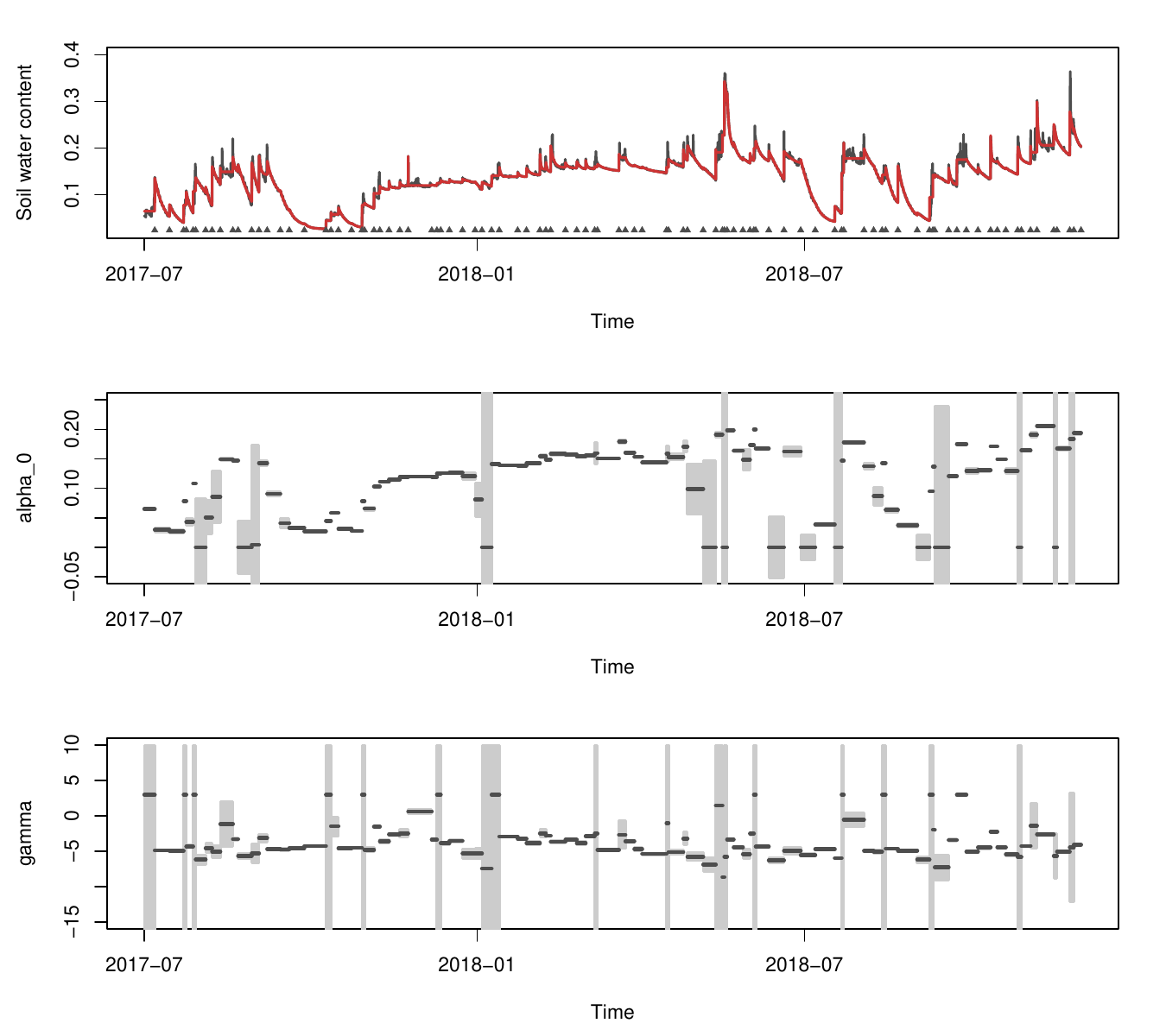}
\captionsetup{labelfont=bf, font=small}
\caption{(top) The soil moisture time series and the estimated changepoints (black triangles) at location 1, field site SERC. (middle) The estimated asymptotic parameter $\alpha_{0}$ over time with a light grey uncertainty band. (bottom) The estimated decay parameter $\gamma$ over time with coloured uncertainty band.}
\label{fig:PELT-fit-SERC}
\end{center}
\end{figure}

\begin{figure}[!htb]
\begin{center}
\includegraphics[width=6in]{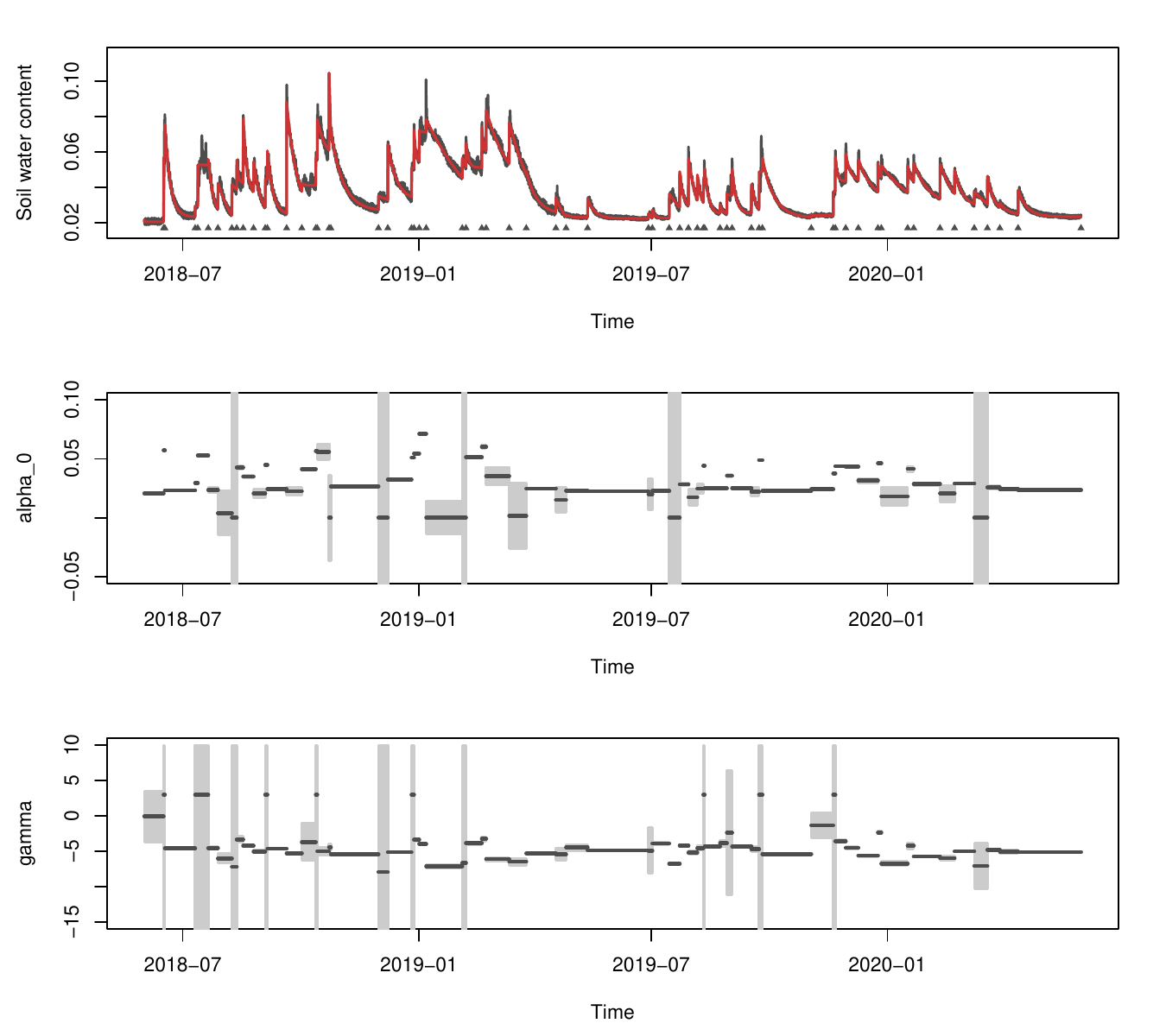}
\captionsetup{labelfont=bf, font=small}
\caption{(top) The soil moisture time series and the estimated changepoints (black triangles) at location 4 in field site SRER. (middle) The estimated asymptotic parameter $\alpha_{0}$ over time with a light grey uncertainty band. (bottom) The estimated decay parameter $\gamma$ over time with coloured uncertainty band. }
\label{fig:PELT-fit-SRER}
\end{center}
\end{figure}

\begin{figure}[!htb]
\begin{center}
\includegraphics[width=6in]{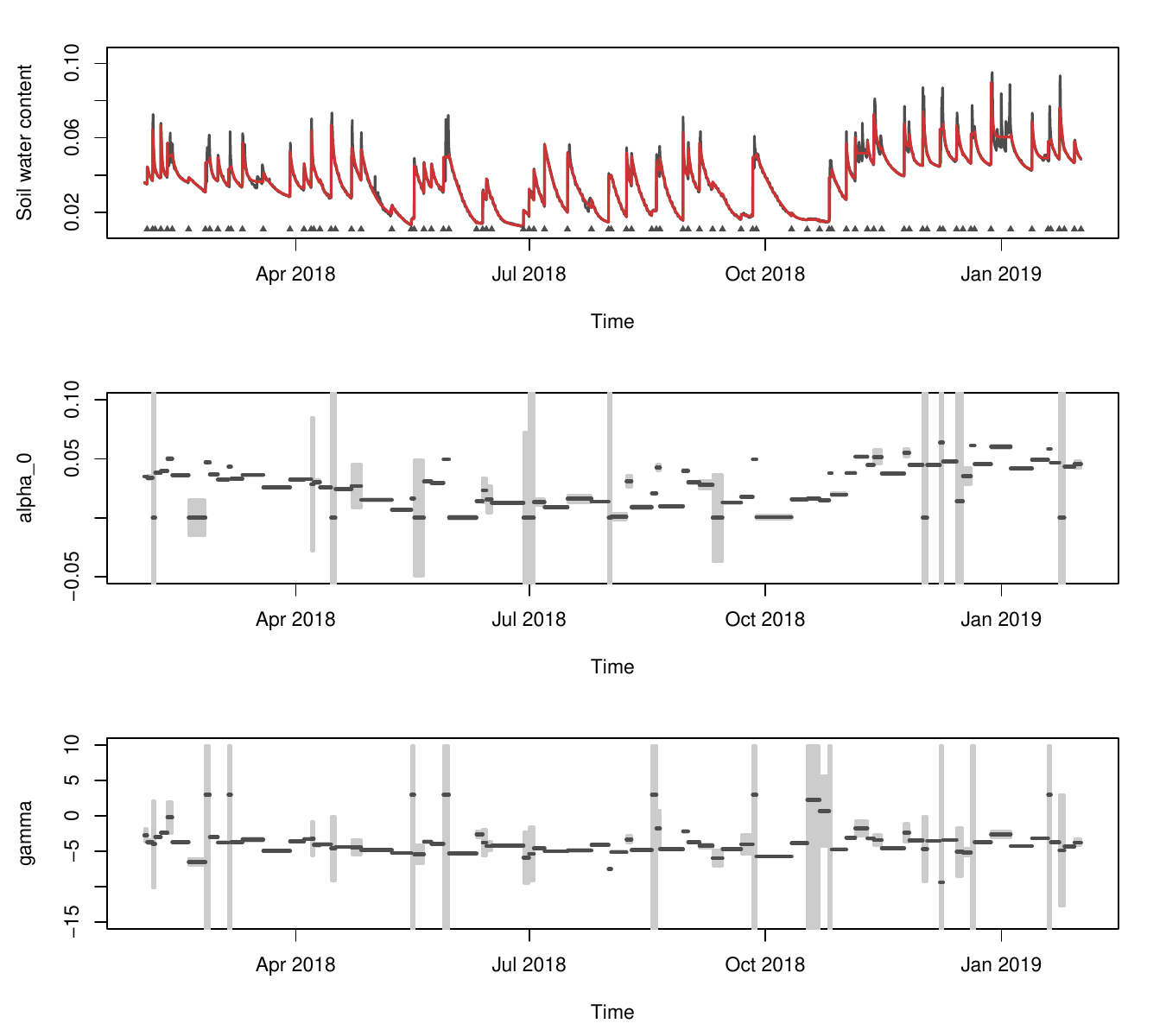}
\captionsetup{labelfont=bf, font=small}
\caption{(top) The soil moisture time series and the estimated changepoints (black triangles) at location 5, field site TALL. (middle) The estimated asymptotic parameter $\alpha_{0}$ over time with a light grey uncertainty band. (bottom) The estimated decay parameter $\gamma$ over time with coloured uncertainty band. }
\label{fig:PELT-fit-TALL}
\end{center}
\end{figure}

The identified changepoints (the black triangles) and the fitted time series (the red curves) are presented in the top panels of Figure \ref{fig:PELT-fit-SERC} to \ref{fig:PELT-fit-TALL} corresponding to field site SERC, SRER and TALL respectively. The estimated asymptotic soil moisture parameter $\alpha_{1}$ and the exponential decay parameter $\gamma$ (the black lines) for each segment along with the uncertainty bands (the light grey bands) are presented in the 2nd and the 3rd panels. To be precise, $\alpha_{1}$ and $\gamma$ are plotted as piecewise constant functions covering the span of the corresponding segments. This helps to visualise the temporal dynamics in the estimated parameters. Due to the lower and upper limits used in the NLS optimisation, the standard errors of parameters at the boundaries tend to be very large. Therefore, the range of the y-axis was decreased, and the very large standard errors are shown as light grey bands stretching from the bottom to the top of the canvas. The e-folding decay parameter $\omega$ (in days) in the soil drydown model (\ref{eqn:drydown}) is also computed from the estimated exponential decay parameter $\gamma$ and the results are given in the figures in the supplemental document. 

All three models achieved a reasonable visual fit, where the red curves in the top panel of Figure \ref{fig:PELT-fit-SERC} to \ref{fig:PELT-fit-TALL} captured the main temporal patterns in the data. Some of the sudden increases are missed due to their short distance to adjacent increases. The lack of fit in some parts of the time series, e.g.\ around July 2018 in site SERC, was associated with the relatively slow increase in the soil moisture which contradicts the assumption of a sudden increase. However, these are not common features. There is no distinctive temporal pattern in the occurrence of the changepoints in this case. 

There is a clear difference in the estimated asymptotic soil moisture $\alpha_{0}$ between the field sites. The asymptotic soil moisture in site SERC is generally higher than that in SRER and TALL (see also the histograms in the supplemental document). This is understandable as site SRER experiences a desert-like climate and site TALL, though humid, has high temperatures year-round. There appear to be some temporal variations in $\alpha_{0}$ as well. For example, for site SERC, the $\alpha_{0}$ during the winter 2018 period behaved slightly differently from the summer period. This suggests that the approach taken in conventional soil drydown modelling where the asymptotic soil moisture is fixed throughout time may not be appropriate. Allowing $\alpha_{0}$ to change over time has the potential to improve the estimation of other parameters in the drydown model.

The differences in the scale of the estimated exponential decay parameter $\gamma$ are less distinctive. The estimated $\gamma$ for site SRER suggests a slightly slower drying rate than the other two sites. This can also be seen from the figures displaying the e-folding decay parameter $\omega$ in the supplemental document. This could be explained by the sparse desert vegetation at site SRER extracting water more slowly or the low unsaturated hydraulic conductivity of the dry soil at SRER, which results in little drainage to lower soil layers. Although there are temporal variations in the estimated decay parameter, there is no clear trend or seasonal pattern in the result here. A longer time series would potentially reveal more interesting features in the dynamics of soil moisture. 

Fundamentally, conventional soil drydown modelling relies on additional hydrological or physical information; in contrast, the proposed method is data-driven. In particular, all parameters of the drydown model are allowed to vary over time. As a result, there will be differences between the parameters estimated using the changepoint method and those from soil drydown literature. Investigating these differences may lead to a better understanding of soil drydown. To summarise, the proposed method provides a different insight into the soil moisture dynamics which is not available using the conventional modelling approach.

\section{Discussion} \label{sec:discussion}
This paper proposed a changepoint-based method to investigate the temporal dynamics in the soil moisture time series. The method aims to identify the structural changes in the form of sudden increases in soil moisture and estimate the parameters characterising the drying process that follows the sudden increase. The method is related to the soil drydown modelling but takes a different approach. It does not rely on the manual identification of soil drydown curves from a soil moisture time series. Instead, it applies a changepoint detection algorithm directly on the soil moisture time series, which automatically identifies the segments representing the exponential decay of soil moisture. The estimation of the soil moisture decay parameters is carried out simultaneously. In addition, the method can be applied to soil moisture time series with little data pre-processing. Thus, when compared to conventional soil drydown modelling, the proposed method has the advantage of easy implementation to a large data set with minimal data preparation. The method also has the flexibility to make use of relevant information, e.g., the precipitation time series data, to improve the segmentation. The simulations and data examples demonstrated the ability of the proposed approach to recover important features of soil moisture drydown. 

Unlike the simulated time series, the real soil moisture time series can display patterns beyond the simple exponential decay. For example, when the soil is saturated during wet seasons or when it is frozen during the winter, the soil moisture time series can show very different patterns from a drydown curve. In practice, it may be sensible to focus on the periods when drydown processes are dominating. Introducing additional information, e.g., temperature or seasonal regimes, may also help to capture different drying patterns. Alternatively, it may be helpful to develop methods that do not rely on the exponential decay assumption and use more flexible models to describe different drying patterns.

Evaluating the uncertainty of the estimated changepoints, though important in some applications, is difficult in a multiple changepoints detection problem. \cite{SpikeCI} proposed a method to compute the confidence intervals for the changepoints in the calcium time series by finding the maximum disturbance that generates the same changepoint. However, the efficient computation of the confidence intervals is tailored to the functional pruning algorithm and therefore is not suitable for use with PELT. Other approaches to quantifying uncertainty in the literature, such as bootstrapped confidence intervals \citep{Bootstrap, BootstrapCpt} and posterior distributions of the changepoint numbers/locations \citep{BayesianCpt, HMMstorm}, do not generalise to the changepoint problem in this paper easily. Therefore, the uncertainty of the identified changepoints is not considered here.

Extending the current method to cover these situations will be a piece of important future work. For example, a useful extension would be to relax the model assumptions and use more flexible models to describe soil moisture dynamics in various scenarios, such as dry and saturated conditions. Potential solutions include re-formatting the problem as a state space model where different states represent different scenarios. Such a model, when estimated within a Bayesian framework, may also provide uncertainty measures to the estimated changepoints. 

Finally, the changepoint method described in section \ref{sec:method} shares some similarities with the so-called shot noise model, which involves a compound Poisson process describing the intensity of the shots and an impulse-response function describing the decay pattern \citep{ShotNoise, NonlinearSN}. In the classical shot noise model, the decay pattern is often modelled as an exponential decay, and it has been used in \cite{ShotNoiseInfiltrate} to estimate the rainfall infiltration, which contributes to the soil moisture dynamics. The shot noise model does not rely on the arrival times of the shots, and hence there is no need to identify the changepoints. However, parameters in the compound Poisson process and the impulse-response function are required. It can be difficult to estimate decay patterns that are changing over time, which is a key feature that the changepoint-based method seeks to expose.

\vspace{2cm}
\textbf{Acknowledgements} \\

The authors gratefully acknowledge the support from the UKRI-funded project Signals in the Soil (Grant No. NE/T012307/1). CN was supported by the Engineering and Physical Sciences Research Council (EPSRC), grant numbers EP/V022636/1, EP/S00159X/1 and  EP/Y028783/1.
 \\

\textbf{Declaration of Competing Interest} \\

The authors declare that they have no known competing financial interests or personal relationships that could have appeared to influence the work reported in this paper. \\

\textbf{Code and data availability} \\

The changepoint detection algorithm and the analysis of the soil moisture time series are implemented in R (version 4.3.1). The code can be accessed from the GitHub repository \url{https://github.com/GMY2018/Changepoint4soil}. 

The data used are publicly available from the United States National Ecological Observation Network (NEON, \url{https://data.neonscience.org/}). The soil moisture time series data can be accessed from \url{https://data.neonscience.org/data-products/DP1.00094.001}. The precipitation data can be accessed from \url{https://data.neonscience.org/data-products/DP1.00006.001}. \\


\end{document}


\maketitle

\section{The modified algorithm when NLS does not converged}

Here we presents the modification of the main algorithm proposed in the main manuscript when the non-linear least square estimation does not converge. 

\begin{algorithm}[h]
\SetKwInOut{Input}{Input}
\SetKwInOut{Output}{Output}

\BlankLine
\textbf{Iterate} \;
\For{$t \leftarrow 2l$ \KwTo $n$}{
1. find all $\tau \in \mathrm{R}_{t}$ that gives finite historical cost $F(\tau)$ and finite cost for the most recent segment $\mathcal{C} (Y_{(\tau+1):t})$, and denote the set of $\tau$ as $\mathrm{P}_{t}$. \\

2. find all $\tau \in \mathrm{R}_{t}$ that gives finite $F(\tau)$ but infinite $\mathcal{C} (Y_{(\tau+1):t})$, and denote the set of $\tau$ as $\mathrm{Q}_{t}$. The set $\mathrm{Q}_{t}$ contains at least one element 0. \\

3. \eIf{$\mathrm{P}_{t} \neq \emptyset$}{follow the iteration in \textbf{Algorithm 1} in the main manuscript}{
(1). find $\tau^{\ast} = \mathrm{arg} \min_{\tau \in \mathrm{Q}_{t}} \left\{ F(\tau) + \lambda \right\}$

(2). update $\mathrm{cpt}(t^{\ast}) = \{ \mathrm{cpt}(\tau^{\ast}), \tau^{\ast} \}$

(3). update $\mathrm{R}_{t+1} = \left\{(t-l+1) \cup \mathrm{Q}_{t} \right\}$ }
}

\caption{Modified iteration} \label{alg:PELT2}
\end{algorithm}

\section{Additional information on the simulation study}

\begin{figure}[!htb]
\begin{center}
\includegraphics[width=5in]{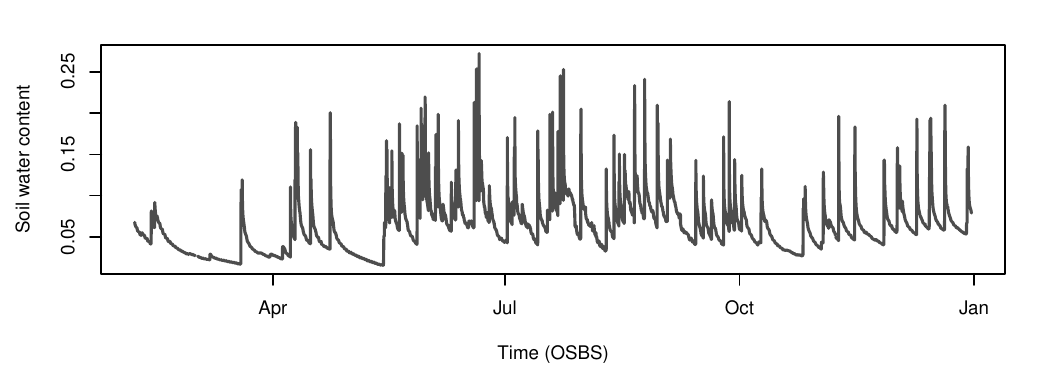}
\includegraphics[width=5in]{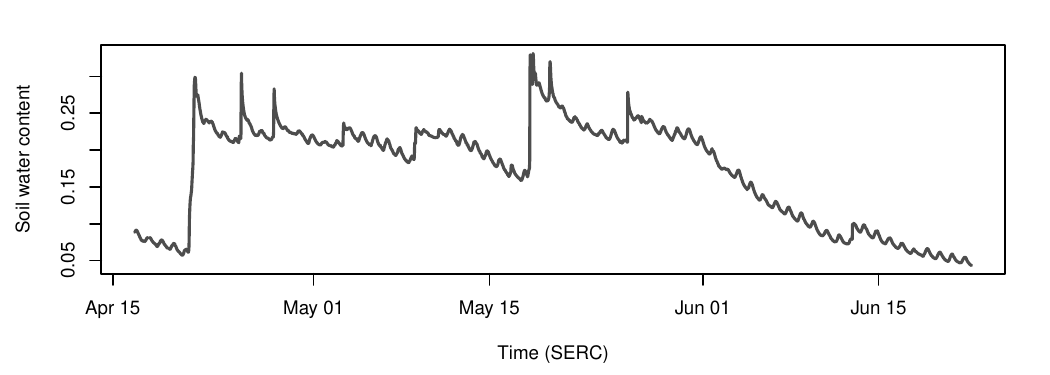}
\captionsetup{labelfont=bf, font=small}
\caption{(top) The soil moisture time series from field site OSBS. (bottom) A close-up of the soil moisture time series from field site SRER.}
\label{fig:data-zoom}
\end{center}
\end{figure}

Figure \ref{fig:data-zoom} presents two soil moisture time series which display interesting features. The upper panel shows a time series with strong seasonal patterns, where the summer months see very frequent increases and falls in soil moisture. The bottom panel shows a time series where smaller increases and falls occur during a long drydown process. Features displayed in these time series motivated the design of the simulation study.

\begin{figure}[!htb]
\begin{center}
\includegraphics[width=6.5in]{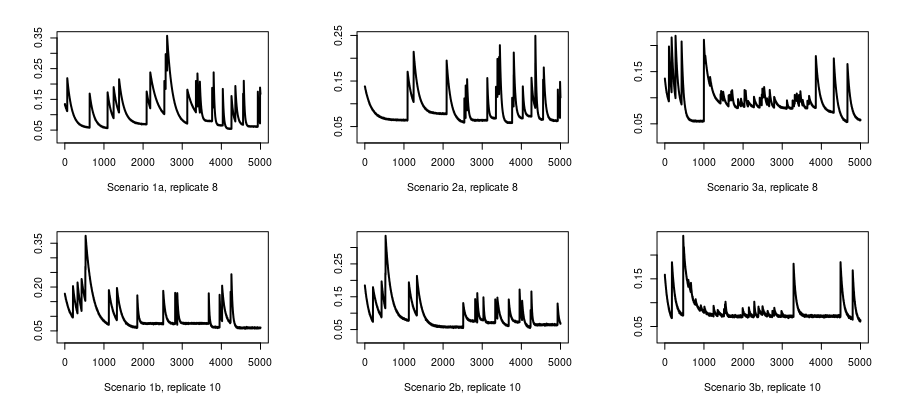}
\captionsetup{labelfont=bf, font=small}
\caption{Examples of the simulated time series from 6 scenarios.}
\label{fig:sim-data}
\end{center}
\end{figure}

Figure \ref{fig:sim-data} shows the simulated time series from six different scenarios described in the main manuscript. The detailed data generation procedure is as follows. 

\begin{enumerate}
\item Generate the changepoints from Poisson processes. In scenario 1a/1b, the changepoints are generated with intensity parameter 0.003. In scenario 2a/2b, the changepoints occur at a lower intensity 0.002 in the first half, and a higher intensity 0.005 in the second half of the time series. Scenario 3a/3b is generated by combining two processes, one over the entire time span $T=5000$ with intensity 0.002, and the other over a period of slow drying with intensity 0.02.

\item Generate the drying rates from uniform distributions. To mimic the seasonal patterns in real data, two uniform distributions U(0.99, 0.995) and U(0.95, 0.99) are used to reflect the slower and faster drying of soil in scenario 1a/1b and 2a/2b. The drying rates of the larger scale increases in scenario 3a/3b are generated from U(0.98, 0.99), and those of the smaller scale increases are generated from U(0.95, 0.99).

\item Generate the increments at the changepoint locations from uniform distributions. The increments in scenario 1a/1b are generated from U(0.1, 0.12). The increments in scenario 2a/2b are generated from U(0.1, 0.12) and U(0.05, 0.1) for periods with fewer and more sudden increases respectively. The increments in scenario 3a/3b are generated from U(0.1, 0.12) and U(0.01, 0.02) for larger scale and smaller scale increases respectively.

\item Generate the asymptotic soil moisture from the uniform distribution U(0.05, 0.08).

\item Generate the Gaussian noise in the data with standard deviation 0.0005 and 0.001, respectively, for the two noise levels. The noise levels are chosen to reflect the sensor precision from NEON data.
\end{enumerate}

\section{An example of penalty learning using NEON data}

Here we present an example of penalty learning using hourly soil moisture time series and hourly precipitation time series from three NEON field sites, SERC, SRER and TALL.

First recall that the loss function used in \cite{PenaltyInterval} and \cite{LearnPenalty} are 
\begin{equation}
    \mathcal{E}(\lambda) = e \left[ \hat{k} (\lambda) \right] = \sum_{(r, a) \in \{ R, A \}} \mathbbm{1} \left( \left| \bm{\tau}^{(\lambda)} \bigcap r \right| \notin a \right) ,
    \label{eqn:PenLoss}
\end{equation}
and 
\begin{equation}
    \mathcal{E}(\lambda) = \mathcal{R}\left( \lambda; Y_{1:n}, \bm{\tau}^{xpt} \right) - \mathcal{R}\left( \lambda; Y_{1:n}, \bm{\tau}^{(\lambda)} \right) \; ,
    \label{eqn:PenRisk}
\end{equation}
respectively. Here, $\hat{k} (\lambda)$ is the estimated number of changepoints given $\lambda$, $\bm{\tau}^{(\lambda)}$ is the optimal set of changepoints given $\lambda$, $\bm{\tau}^{xpt}$ is the experts' annotated changepoints as in \cite{LearnPenalty}, $R$ is a set of regions on the time axis and $A$ is a set of annotations on the possible number of changepoints within each region in set $R$ as in \cite{PenaltyInterval}.

To create the experts' annotations as in \cite{PenaltyInterval}, first define the regions of annotations by dividing the time axis into intervals and discounting any regions with no precipitation data. Here we divide the time axis into regions of 10 days for illustration purpose. More careful design will be required for better learning. Then substantial rainfall instances are first converted into binary data, i.e., $X_{t} = 1$ if the precipitation at time $t$ is above certain threshold, and $X_{t} = 0$ otherwise. As the annotation can be a known number (including 0) of changepoints or a range of numbers, e.g., $a_{1} = \{ 0 \}, a_{2} = \{ 1, 2, 3 \}, a_{3} = \{ 1 \}, \cdots$, we apply two different thresholds (0.5 and 1 milliliter respectively) to the rainfall data to create the ranges of possible numbers of changepoints. Intuitively, a higher threshold will result in a smaller count in each region and a lower threshold will give a larger count. Using these counts, we can create annotations $a_{j}$ for region $r_{j}$, for $j = 1, 2, \cdots, |R|$. Finally, the loss function (\ref{eqn:PenLoss}) can be calculated by comparing the estimated changepoints under penalty $\lambda$ to the counts of changepoints in each region. The penalty parameters that minimises the loss function would be appropriate choices that balance the optimal segmentation based on soil moisture data only and the information from the rainfall data.

To create the experts' labelling as in \cite{LearnPenalty}, rainfall instances grater than 1 milliliter are converted into binary data. The collection of time points $\tau$ with $X_{\tau} = 1$ then becomes the experts' labelling $\bm{\tau}^{xpt}$. Given the penalty parameter $\lambda$, the risk of following the experts' labelling $\mathcal{R}\left( \lambda; Y_{1:n}, \bm{\tau}^{xpt} \right)$ can be calculated as the overall cost of the segments with changepoints $\bm{\tau}^{xpt}$, and the risk of the optimal segmentation given $\lambda$, $\mathcal{R}\left( \lambda; Y_{1:n}, \bm{\tau}^{(\lambda)} \right)$, can be obtained by running the proposed changepoint detection algorithm. The excess penalised risk (\ref{eqn:PenRisk}) is then evaluated on a sequence of penalty parameters $\lambda$. The range of penalty parameters that minimises the excess penalised risk would be an appropriate choice for the changepoint detection problem. If there are missing gaps in the precipitation time series, the annotation will be incomplete. In such case, the cost corresponding to the missing gaps needs to be discounted.

\begin{figure}[!htb]
\begin{center}
\includegraphics[width=5.8in]{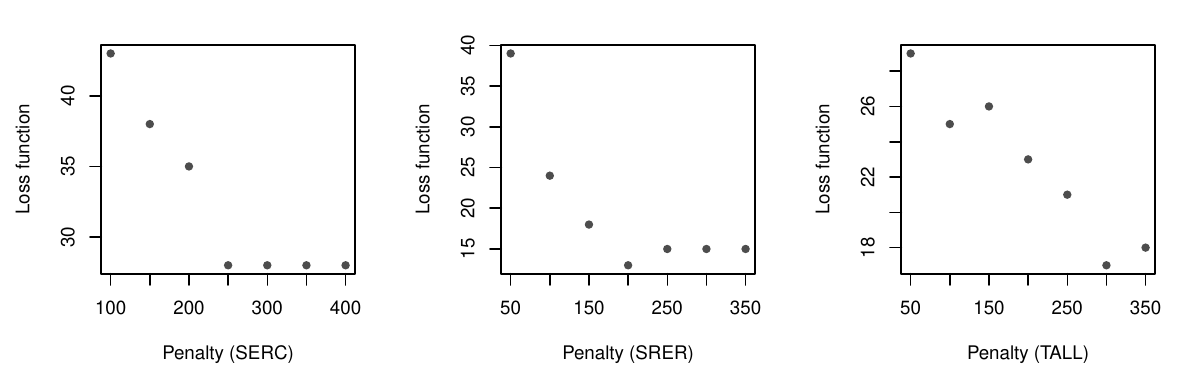}
\includegraphics[width=5.8in]{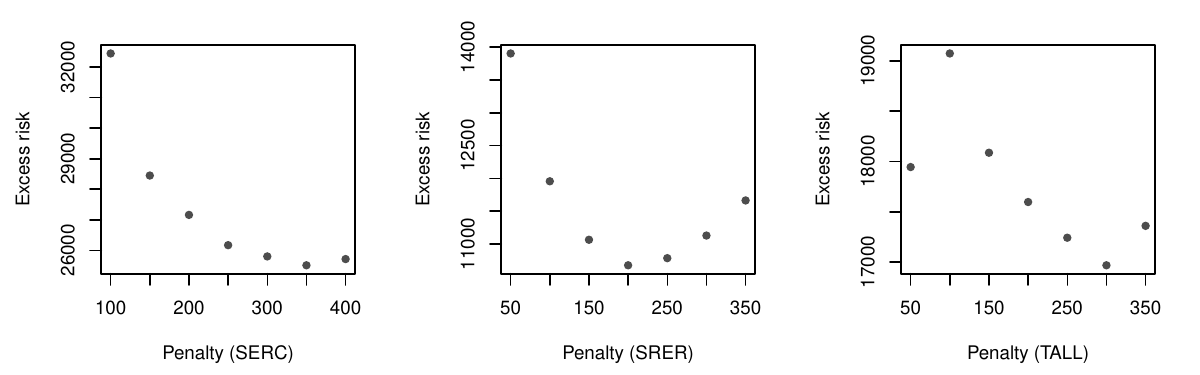}
\captionsetup{labelfont=bf, font=small}
\caption{(Top) Examples of loss function for comparing the experts' annotations (i.e., the possible number of changepoints within each region) and the PELT segmentation against the penalty parameter, from NEON field site SERC, SRER and TALL respectively. (Bottom) Examples of excess penalised risk between the experts' labelling (i.e., large rainfall instances) and the PELT segmentation against the penalty parameter, from NEON field site SERC, SRER and TALL respectively.}
\label{fig:ExRisk}
\end{center}
\end{figure}

\begin{figure}[!htb]
\begin{center}
\includegraphics[width=5.5in]{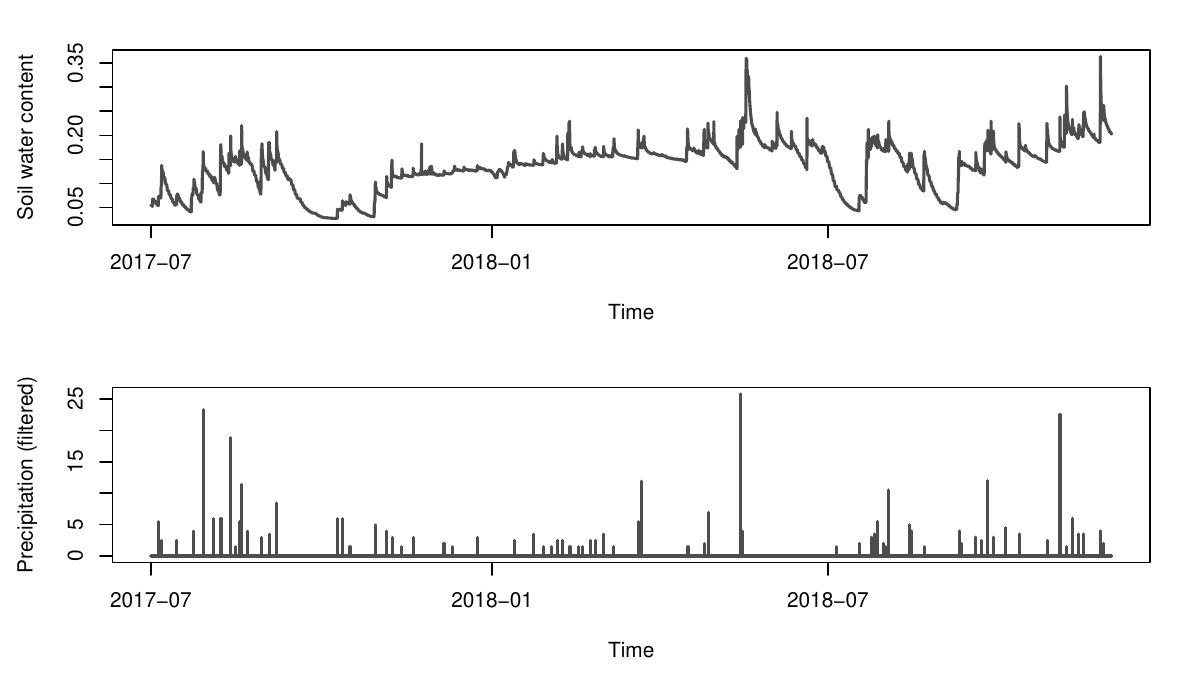}
\captionsetup{labelfont=bf, font=small}
\caption{The soil moisture time series from NEON site SERC and the precipitation time series after filtering out the rainfall instances $\leq$ 1 milliliter.}
\label{fig:sm-rain-serc}
\end{center}
\end{figure}

\begin{figure}[!htb]
\begin{center}
\includegraphics[width=5.5in]{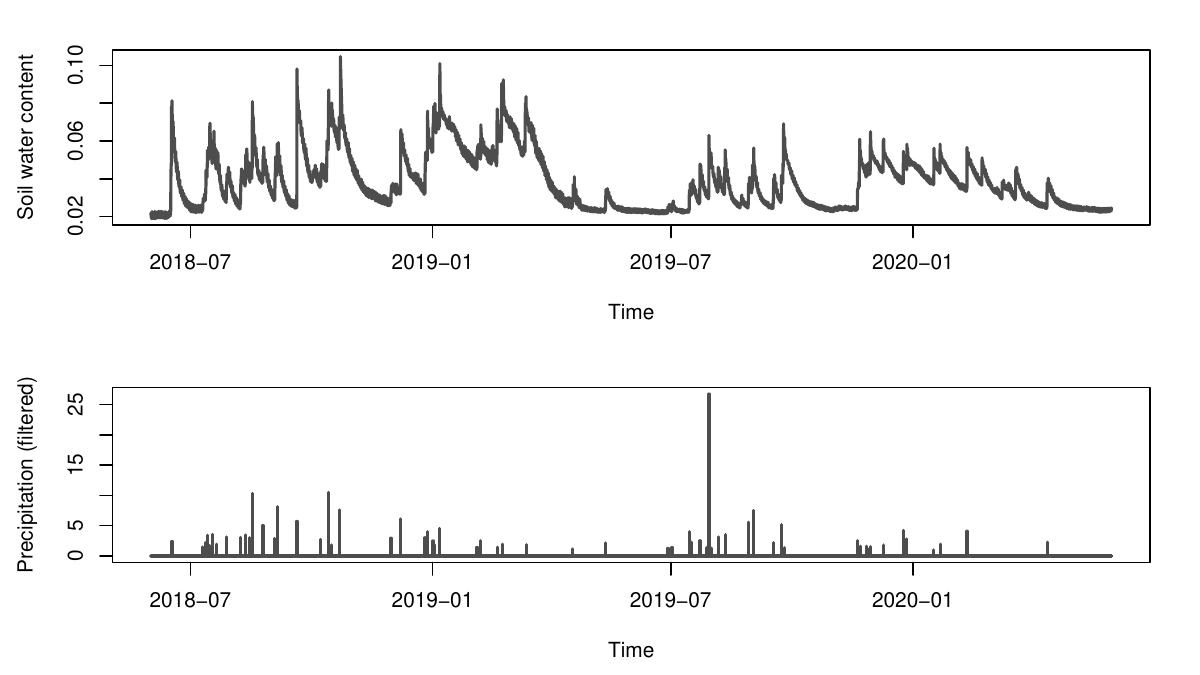}
\captionsetup{labelfont=bf, font=small}
\caption{The soil moisture time series from NEON site SRER and the precipitation time series after filtering out the rainfall instances $\leq$ 1 milliliter.}
\label{fig:sm-rain-srer}
\end{center}
\end{figure}

\begin{figure}[!htb]
\begin{center}
\includegraphics[width=5.5in]{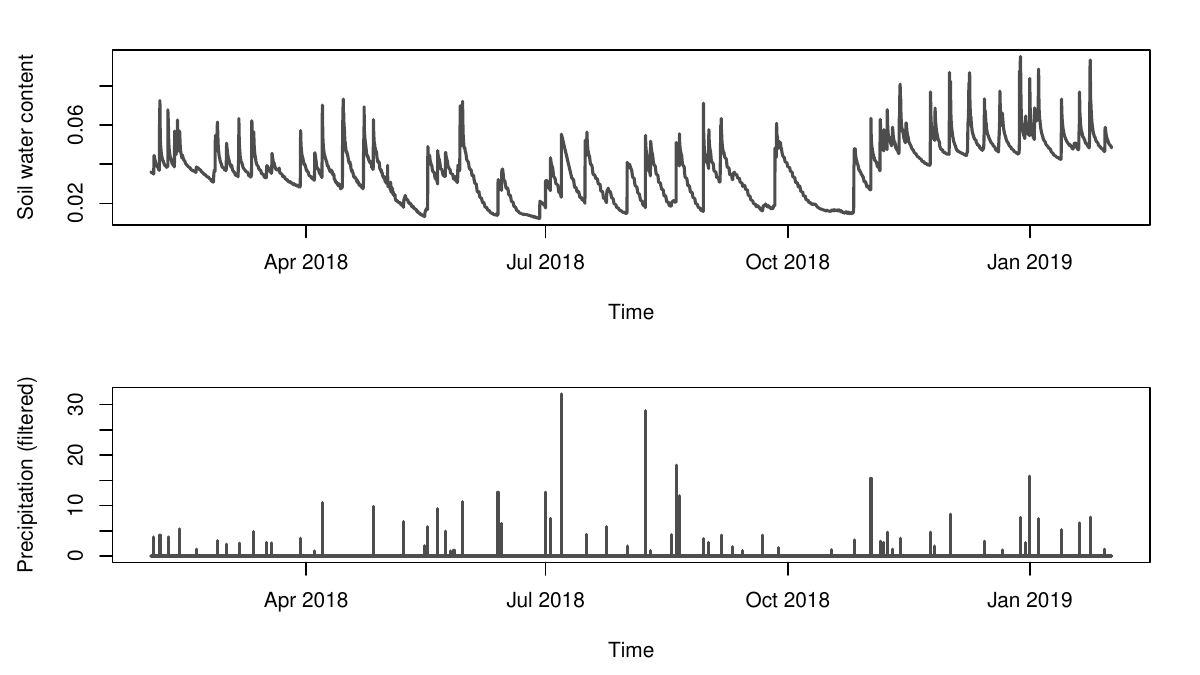}
\captionsetup{labelfont=bf, font=small}
\caption{The soil moisture time series from NEON site TALL and the precipitation time series after filtering out the rainfall instances $\leq$ 1 milliliter.}
\label{fig:sm-rain-tall}
\end{center}
\end{figure}

The results from applying the method in \cite{PenaltyInterval} to the soil moisture time series from the three field sites, SERC, SRER and TALL, are shown in top panels of Figure \ref{fig:ExRisk}. The results from applying the ALPIN method in \cite{LearnPenalty} are shown in bottom panels of Figure \ref{fig:ExRisk}. The results from the two penalty learning methods appear to be consistent. In particular, the plots suggest a relatively large penalty ($\lambda > 250$ or $\lambda > 300$) for soil moisture time series from SERC, a penalty around 200 for SRER and a penalty of 300 for TALL. For information, Figures \ref{fig:sm-rain-serc} to \ref{fig:sm-rain-tall} show the correspondence between bulk precipitation greater than 1 milliliter and the peaks in the soil moisture time series from the three field sites. 

\section{Additional information on the application}

\begin{figure}[!htb]
\begin{center}
\includegraphics[width=6in]{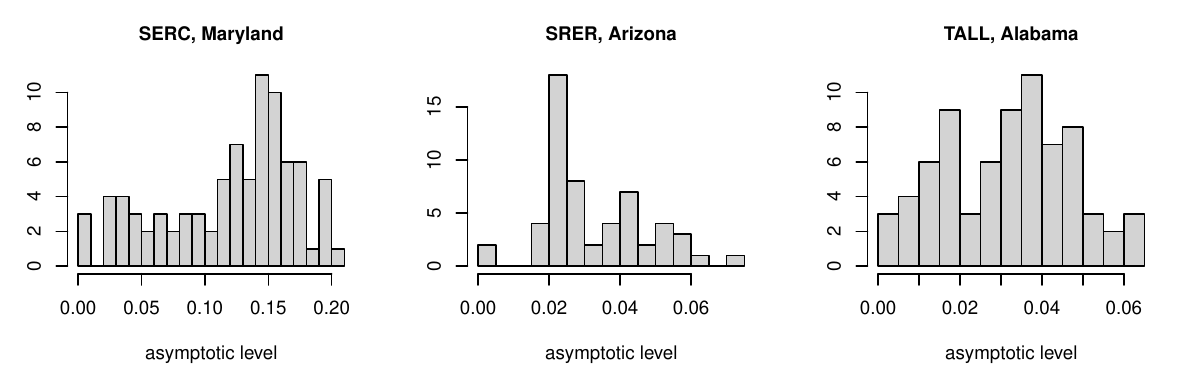}
\includegraphics[width=6in]{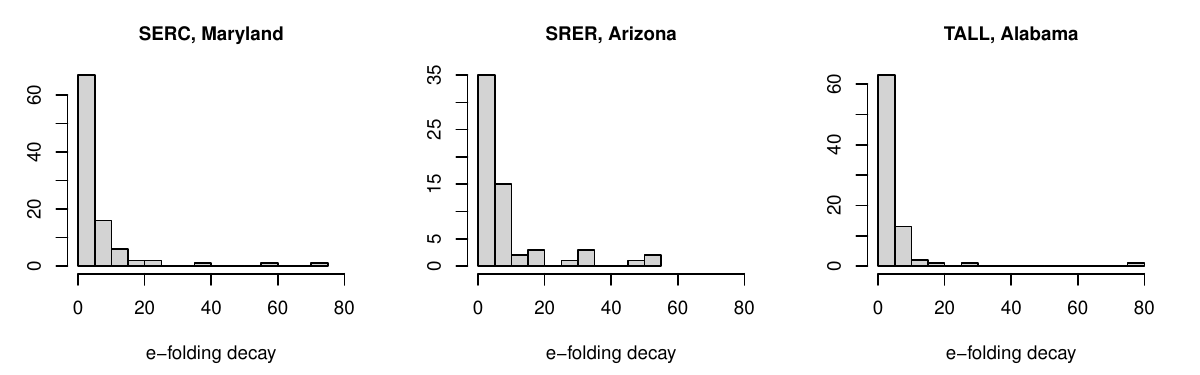}
\captionsetup{labelfont=bf, font=small}
\caption{Histograms of the estimated asymptotic level and the e-folding decay parameter computed from the estimated decay parameters for SERC, SRER and TALL.}
\label{fig:hist-params}
\end{center}
\end{figure}

\begin{figure}[!htb]
\begin{center}
\includegraphics[width=5.8in]{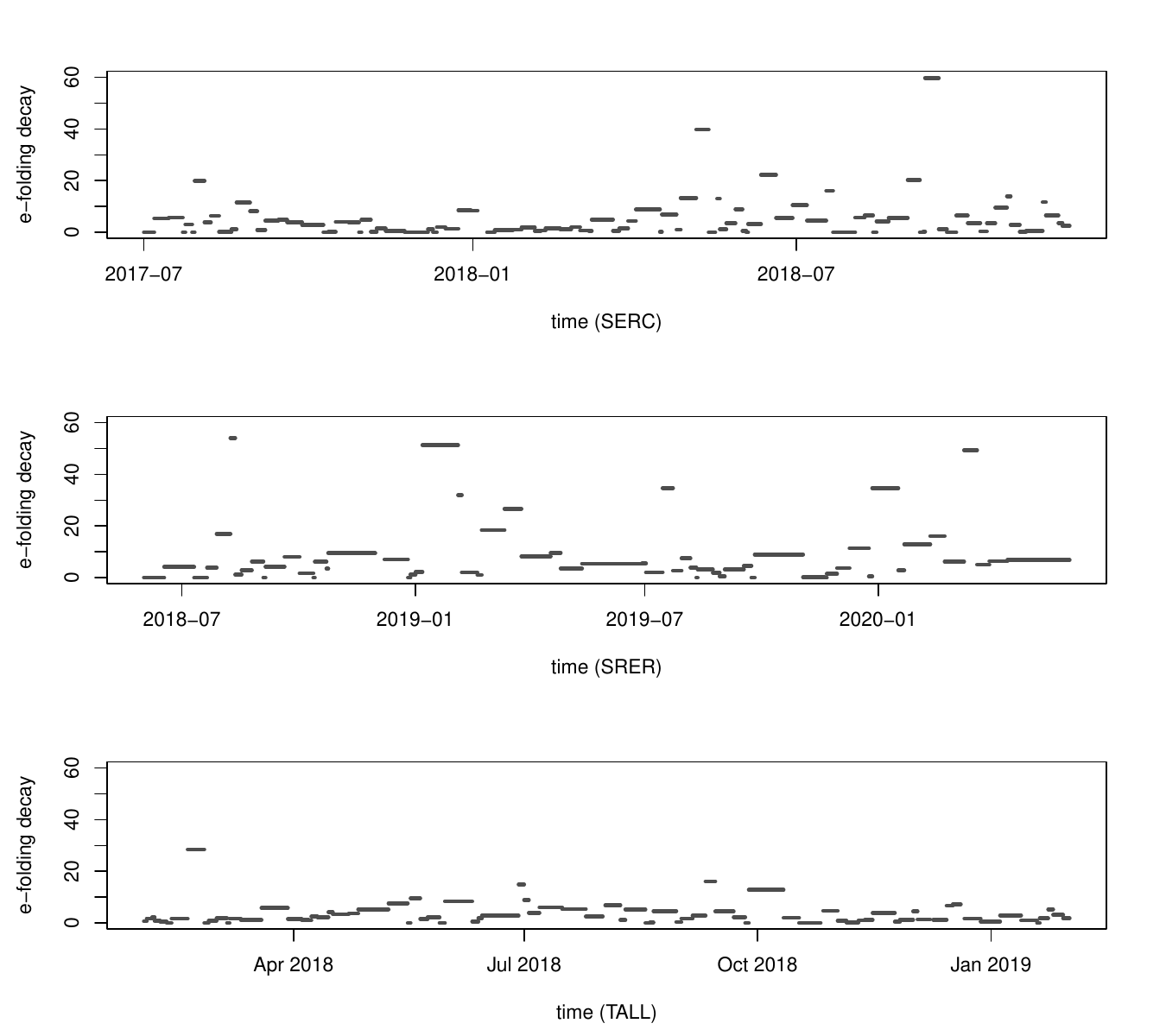}
\captionsetup{labelfont=bf, font=small}
\caption{The e-folding decay parameter computed from the estimated decay parameters plotted over time for SERC (top), SRER (middle) and TALL (bottom). }
\label{fig:efolding-decay-ts}
\end{center}
\end{figure}

Figure \ref{fig:hist-params} shows the histograms of the estimated asymptotic soil moisture $\alpha_{0}$ and the estimated e-folding decay parameter $\omega$, which is calculated from the decay parameter $\gamma$, from the three field sites. Figure \ref{fig:efolding-decay-ts} shows the e-folding decay parameter plotted over time. There are clear differences in the asymptotic level parameters, but the differences in the decay parameters (whether it is in $\gamma$ scale or e-folding scale) are not obvious. 

